\documentclass[10pt,twocolumn]{article}
\RequirePackage[letterpaper]{geometry}

\usepackage{fullpage}
\RequirePackage{times}
\usepackage{xcolor}
\definecolor{darkgreen}{rgb}{0,0.4,0}
\definecolor{darkorange}{rgb}{1.0,0.4,0.0}

\usepackage[normalem]{ulem}
\RequirePackage[breaklinks=true,colorlinks=true,citecolor=black,linkcolor=darkgreen,urlcolor=blue]{hyperref}
\usepackage[nolist,nohyperlinks]{acronym}
\RequirePackage{amsmath} 
\RequirePackage{amssymb} 
\RequirePackage{bm} 
\RequirePackage{mathptmx}
\usepackage{subfigure}    
\usepackage{wrapfig}    
\usepackage{graphicx}   
\usepackage[export]{adjustbox} 
\usepackage{mwe} 
\usepackage[small,bf]{caption} 
\usepackage[compact]{titlesec} 

\usepackage[hang,flushmargin]{footmisc} 

\usepackage{tabularx}

\newcommand{\ignore}[1]{}

\titleformat{\section}{\normalfont\Large\bfseries}{\thesection}{0.5em}{}
\titlespacing{\section}{0pt}{*0.5}{*0.5} 
\titleformat{\subsection}{\normalfont\large\bfseries}{\thesubsection}{0.5em}{}
\titlespacing{\subsection}{0pt}{*0.5}{*0.5}
\titleformat{\subsubsection}{\normalfont\bfseries}{\thesubsubsection}{0.5em}{}
\titlespacing{\subsubsection}{0pt}{*0.3}{*0.3}
\titleformat{\paragraph}{\normalfont\bfseries}{\theparagraph}{0.5em}{}
\titlespacing{\paragraph}{0em}{0em}{0.0em}
\titleformat{\subparagraph}{\normalfont\bfseries}{\thesubparagraph}{0.5em}{}
\titlespacing{\subparagraph}{0pt}{0pt}{0em}
\titlespacing*{\subparagraph}{0pt}{0pt}{0em}

\usepackage{enumitem}
\RequirePackage[compress,capitalise,noabbrev]{cleveref} 
\crefformat{equation}{(#2#1#3)} 
\crefname{section}{Sec.\/}{Secs.\/}
\Crefname{section}{Sec.\/}{Secs.\/}
\crefname{subsection}{Sec.\/}{Secs.\/}
\Crefname{subsection}{Sec.\/}{Secs.\/}
\crefname{subsubsubsection}{Sec.\/}{Secs.\/}
\Crefname{subsubsubsection}{Sec.\/}{Secs.\/}
\crefformat{paragraph}{Sec.\ #2#1#3} 
\crefformat{subparagraph}{Sec.\ #2#1#3} 
\crefname{figure}{Fig.\/}{Figs.\/}
\Crefname{figure}{Fig.\/}{Figs.\/}

\newcommand{\greenHL}[1]{{\color{darkgreen}\bf #1}}
\newcommand{\redHL}[1]{{\color{red}\bf #1}}
\newcommand{\jnote}[1]{{\color{blue}#1}}

\newcommand{\jsumm}[1]{{\hfil{\scriptsize\color{blue}\uline{#1}}\hfil}\\}

\definecolor{mylightgrey}{rgb}{0.6,0.6,0.6}
\definecolor{mydarkgrey}{rgb}{0.3,0.3,0.3}
\definecolor{mybrown}{rgb}{0.6,0.2,0}
\definecolor{mylightgreen}{rgb}{0,1,0}
\definecolor{mydarkgreen}{rgb}{0,0.5,0}
\definecolor{MyDarkGreen}{rgb}{0.0,0.4,0.0} 
\definecolor{myorange}{rgb}{1,0.4,0}
\definecolor{mypurple}{rgb}{0.62,0.0,0.35}
\definecolor{MyDarkRed}{rgb}{0.8,0.0,0.0} 
\definecolor{shockingpink}{rgb}{0.988,0.059,0.753}

\newcommand{\rmwords}[1]{{\color{mypurple} #1}}
\newcommand{\rmtolookat}[1]{{\color{shockingpink} #1}}

\newcommand{\eg}{{\it e.g.\/}} 
\newcommand{\ie}{{\it i.e.\/}} 
\newcommand{\etc}{{\it etc.\/}} 
\newcommand{\wrt}{{\it wrt}}

\newcommand{\etal}{{\it et al.}} 
\newcommand{\vs}{{\it vs.\/}}
\newcommand{\longemdash}{\rule[.5ex]{1.3em}{0.5pt}}

\newcommand{\be}[1]{\begin{equation}\label{#1}}
\newcommand{\ee}{\end{equation}}

\newcommand{\putDOI}[1]{}

\renewcommand{\baselinestretch}{0.88}\selectfont

\let\tmpignore=\ignore
\let\newRedHL=\redHL
\let\keepRedHL=\redHL

\let\jnote=\ignore
\let\redHL=\ignore
\let\jsumm=\ignore
\let\keepRedHL=\ignore

\begin{document}

\abovedisplayshortskip=0.3em
\belowdisplayshortskip=0.2em
\abovedisplayskip=0.3em
\belowdisplayskip=0.2em
\arraycolsep 0pt 

\begin{acronym} 
    \acro{TITLE}[\acs{FPIM}]{\acl{FPIM}s for Solving SAT} 
    \acro{QuICC}[QuICC]{Quantum-Inspired Classical Computing}
    \acro{cSoA}[cSoA]{Current State-Of-the-Art}
    \acro{OIM}[OIM]{Oscillator Ising Machine}
    \acro{CIM}[CIM]{Coherent Ising Machine}
    \acro{CIMpp}[\textbf{CIM++}]{Improved Coherent Ising Machine (CMOS IC)}
    \acro{CO}[CO]{Combinatorial Optimization}
    \acro{BRIM}[BRIM]{Bistable Resistive Ising Machine}
    \acro{BLIM}[BLIM]{Bistable Latch Ising Machine}
    \acro{CDR}[BCR]{Bits of Coupling Resolution}
    \acro{BCR}[BCR]{Bits of Coupling Resolution}
    \acro{FPGA}[FPGA]{Field-Programmable Gate Array}
    \acro{IM}[IM]{Ising Machine}
    \acro{FPIM}[\textbf{FPIM}]{Field-Programmable Ising Machine}
    \acro{FPOIM}[\textbf{FPOIM}]{Field-Programmable Oscillator Ising Machine}
    \acro{SAT}[SAT]{Boolean Satisfiability}
    \acro{QuICCSAT}[\textbf{QuICCSAT}]{Scalable, sparse, low-BCR SAT-to-Ising mapping}
    \acro{OIMLI}[\textbf{OIMLI}]{Oscillator Ising Machine for Learning and Inference}
    \acro{ImLI}[\textbf{ImLII}]{Image Learning and Inference via Ising}
    \acro{DEaSP}[\textbf{Dense-as-Sparse}]{Approximating Dense Connectivity with Sparse Resistive Networks}
    \acro{DivConqBCR}[\textbf{DivConq-BCR}]{Divide and Conquer by Bits of Coupling Resolution}
    \acro{DivConqSize}[\textbf{DivConq-Size}]{Divide and Conquer by Problem Size}
    \acro{MILP}[MILP]{Mixed Integer Linear Programming}
    \acro{MFMC}[MFMC]{Maximum Fault Minimum Cardinality}
    \acro{MIMOMLE}[MIMO-MLE]{Maximum Likelihood Estimation for Wireless MIMO Decoding}
    \acro{MIMOMLEI}[\textbf{MIMO-MLE-Ising}]{\acs{MIMOMLE} to Ising}
    \acro{MILPI}[\textbf{\acs{MILP}-Ising}]{\acs{MILP} to Ising}
    \acro{MFMCI}[\textbf{\acs{MFMC}-Ising}]{\acs{MFMC} to Ising}
    \acro{xIMSIM}[\textbf{xIMSIM}]{Analog Hardware Performance Models for \acs{TITLE}}
    \acro{SER}[SER]{Symbol Error Rate}
    \acro{BER}[BER]{Bit Error Rate}
    \acro{BM}[BM]{Boltzmann Machine}
    \acro{RBM}[RBM]{Restricted Boltzmann Machine}
    \acro{PROPRIETARY}[{\small\textbf{PROPRIETARY}}]{Proprietary Content}
\end{acronym}

\title{\vskip-2em \acs{TITLE}: \acl{TITLE}}
\author{Thomas Jagielski\thanks{Yale University},\/~\ Rajit Manohar\footnotemark[1]\/~\ and Jaijeet Roychowdhury\thanks{University of California, Berkeley.}}
\date{}
  
\maketitle

\thispagestyle{plain}
\renewcommand{\thesection}{\Roman{section}}
\renewcommand{\thesubsection}{\thesection.\Alph{subsection}}
\newcommand{\BadGuys}{AhChMoKi2021JSSCringOscIMshort,MaBaTrCaJoSh2020NCOMMSrelaxationOscIsingCMOS,DuKhAsPaScToRaDaVO2eightSpinIM}

\begin{abstract}
On-chip analog Ising Machines (IMs) are a promising means to solve difficult combinatorial optimization problems.
For scalable on-chip realizations to be practical, 1) the problem should map scalably to Ising form, 2) interconnectivity between spins should be sparse, 3) the number of bits of coupling resolution (\acs{BCR}) needed for programming interconnection weights should be small, and 4) the chip should be capable of solving problems with different connection topologies.
We explore these issues for the SATisfiability problem and devise \acs{TITLE}, a reconfigurable on-chip analog Ising machine scheme well suited for SAT.
To map SAT problems onto \acs{FPIM}s, we leverage Boolean logic synthesis as a first step, but replace synthesized logic gates with Ising equivalent circuits whose analog dynamics solve SAT by minimizing the Ising Hamiltonian.
\jnote{Our reconfigurable interconnections between Ising spins, thereby removing limitations in prior approaches to fixed regular topologies.}
We apply our approach to 2000 benchmark problems from SATLIB, demonstrating excellent scaling, together with low sparsity and low BCR that are independent of problem scale.
Placement/routing reveals a very feasible requirement of less than 10 routing tracks to implement all the benchmarks, translating to an area requirement of about $10$mm$^2$ for a programmable 1000-spin \acs{FPIM} in 65nm technology.
\jnote{
These results indicate that large-scale SAT-solving Ising machine chips will be practical in the near term.
}
\end{abstract}


\tmpignore{
\section{Planning}
\label{sec:planning}
\begin{enumerate}[leftmargin=0.9em,itemsep=-0.25em,topsep=0em,itemindent=-0.0em,labelwidth=0.3em]%
\item Mapping SAT to a logic gate network 
\item Mapping each gate to an Ising equivalent (differences if any from Su, ?Dattani?) \redHL{pretty much all done by Su}
     \begin{enumerate}[leftmargin=1.6em,itemsep=-0.25em,topsep=-0.2em,itemindent=-0.0em,labelwidth=0.3em]%
     \item Proving that min of each Ising Eq gate $\Leftrightarrow$ Gate logic satisfied
     \item mapping non-minimum gates to number of bit errors \redHL{pointed out by Lei He}
     \end{enumerate}
\item Proving that the min.\ H of any composition of Ising eqvt.\ gates corresponds to the same composition of gate level logic being satisfied.
     \begin{enumerate}[leftmargin=1.6em,itemsep=-0.25em,topsep=-0.2em,itemindent=-0.0em,labelwidth=0.3em]%
     \item Note the big difference: gate-level has directionality, Ising doesn't
           \begin{enumerate}[leftmargin=1.0em,itemsep=-0.25em,topsep=-0.2em,itemindent=-0.0em,labelwidth=0.3em]%
           \item Hence can set outputs (?intermediate nodes stuck at for fault analysis?) and let H.\ minimization figure out the correct logic\longemdash\/this is how SAT formulated and solved as Ising.
           \end{enumerate}
     \end{enumerate}
\item Differences from Su
     \begin{enumerate}[leftmargin=1.6em,itemsep=-0.25em,topsep=-0.2em,itemindent=-0.0em,labelwidth=0.3em]%
     \item 3 input gates; XOR with extra node; comprehensive table of gates and Ising equivalents; linalg approach, bit-error related penalty.
     \end{enumerate}
\item Postprocessing/connectivity options and impact on BCR/density \greenHL{some key contribs are here}
     \begin{enumerate}[leftmargin=1.6em,itemsep=-0.25em,topsep=-0.2em,itemindent=-0.0em,labelwidth=0.3em]%
     \item gate library exploration
     \item min.\ no.\ of spins option: dense, higher BCR (examples to quantify)
     \item logic synthesis to contain fanout (more generally, for routability); will require inverters $\rightarrow$ low BCR, sparse (examples to quantify, Rajit already has?). \redHL{Note: Su already uses ABC}
     \item impact of the above on difficulty of solution? using tabu/SA?
     \item also try OIM, try to tweak for good results.
     \end{enumerate}
\item FPIM \greenHL{main contrib}
     \begin{enumerate}[leftmargin=1.6em,itemsep=-0.25em,topsep=-0.2em,itemindent=-0.0em,labelwidth=0.3em]%
     \item Su is placing gates (?); we are placing spins/LUTs
     \item example runs (SA, OIM) on actually-placed
     \end{enumerate}
\end{enumerate}
}

\section{Introduction}
\label{sec:intro}

In recent years, Ising Machines (\acs{IM}s) have gained traction as a viable approach for solving hard combinatorial optimization (\acs{CO}) problems.
\acs{IM}s use specialized hardware, typically based on quantum, probabilistic or analog physics/mathematics, for attacking \acs{CO} problems reformulated in so-called Ising form.
Though \acs{IM}s were initially inspired by quantum computers \cite{JohnsonEtAl2011NatureQuantumAnnealingWithManufacturedSpinsShort,wang2013coherentShort}, a class of analog \acs{IM}s that operate on purely classical principles has arisen.\ignore{\footnote{Indeed, the actual operation of practical systems implementing \acs{CIM} \cite{inagaki2016ScienceIsing2000compact} is widely believed to be purely classical.}}
Such analog \acs{IM}s are able to solve many standard \acs{CO} problems competitively \vs\ quantum \acs{IM}s,\ignore{\footnote{although D-Wave's quantum annealers have been shown to feature quantum advantage \cite{ronnow2014definingShort} on select problems, such as quantum magnetics simulations \cite{KingEtAl2021NatureCommDWaveScalingAdvantageShort}.}} while typically being far simpler, smaller and cheaper.
Particularly promising are analog Ising solvers that can be fabricated on chip, \eg, in standard CMOS technologies\longemdash\/examples include \ignore{hardware simulated annealers \cite{yamaoka2016IsingCMOSshort,AramonEtAl2019FoPDigitalAnnealer},} \acs{OIM} \cite{WaWuNoRoOIMnatComputing2021}, \acs{BRIM}/\acs{BLIM} \cite{AfZhVeIgHu2021HPCABRIMshort,RoUCNC2021BLIMcompact} and related approaches \cite{AhChMoKi2021JSSCringOscIMshort}.\footnote{In analog Ising solvers, circuits such as oscillators or analog latches function as Ising ``spins''. 
  Complex analog dynamics engendered by physical connections between the spins underlie their ability to solve CO problems.
  A very different approach for solving Ising problems is digital emulation, \ie, solving mathematical models of analog Ising machines using fixed- or floating-point numerical methods implemented as custom ICs or on standard FPGAs (\eg, \cite{TaYaGoSimulatedBifurcationMultiChip2021,TaYaGoSimulatedBifurcationFPGA2019}). 
  This work is not directly relevant to digital emulation approaches.}
The potential benefits of such on-chip analog \acs{IM}s over quantum/optical ones include reductions in physical size, energy consumption and cost of many orders of magnitude, as well as feasible scaling to large problem sizes.

However, delivering the above-noted promise for practical CO problems requires careful consideration of the scalability of problem mapping to Ising form and the type of connectivity required (dense or sparse).
For example, the number of Ising spins needed for the Travelling Salesman Problem (TSP) grows quadratically \wrt\ problem size \cite{lucas2014ising}. \redHL{CHECK quadratic.}
Mapping the MU-MIMO detection problem does not lead to more spins \cite{kim2019leveragingShort,RoWaSr2021MUMIMOdetectionRSpreprintCompact}, but all-to-all connectivity is required, presenting an on-chip implementation challenge \cite{SaRoDenseAsSparseICCAD2022}.
Another key issue for on-chip realization is configuring different connectivity patterns to address different problems.
Previous work has relied on re-mapping the Ising problem onto a larger Ising fabric with a fixed sparse connection topology, such as Chimera, Pegasus or King's graphs \cite{BoothbyEtAlChimeraPegasusMapping2020Short,AhChMoKi2021JSSCringOscIMshort,SuTuHeDAC2016SATmappingForDWaveShort}. 
Re-mapping typically increases Ising problem size greatly \cite{SugieEtAl2021minorEmbeddingLargeKingsGraphShort}, severely limiting the sizes of CO problems that can be solved with a fixed number of spins on an IC.

To realize the potential benefits of analog Ising solvers, a CO problem class must have an Ising mapping that does not explode with problem size and also satisfies other constraints tied to realistic on-chip implementation.
In this paper, we show that SAT problems can be mapped to a practical reconfigurable Ising solver architecture, dubbed \acs{FPIM}, that we propose.
\acs{FPIM} is not tied to a specific analog Ising scheme; it can use spins based on, \eg, oscillators \cite{WaWuRoDAC2019OIMlateBreaking}, ZIV diodes \cite{AfZhVeIgHu2021HPCABRIMshort}, CMOS latches \cite{RoUCNC2021BLIMcompact}, \etc.
While the design of \acs{FPIM}s leverages ideas from FPGAs, there are key differences between the two.
In \acs{FPIM}s, \textit{analog} spin circuits (not, \eg, digital LUTs) are coupled bi-directionally using \textit{programmable resistive connections} that implement Ising weights.\footnote{\ie, there is no concept of directional signal flow, unlike in digital circuits.}
Analog operation enables an \acs{FPIM} to solve a CO problem using mechanisms completely different from the digital computations that FPGAs accelerate.

Our core contribution is a novel, effective way to design practical \acs{FPIM}s for SAT problems.
First, we employ logic synthesis tools (\texttt{yosys} \cite{wolf2013yosysShort}, \texttt{ABC} \cite{BrMiCAVABC2010Short}) to obtain gate-level circuits for the SAT functions to be solved.
Then, we use a SAT-to-Ising mapping procedure that replaces each logic gate with an ``Ising equivalent'' consisting of a few spin units with weighted couplings.\footnote{Ising equivalents are defined at the Ising graph level; they can be implemented using any circuit level analog spin scheme.}
Analog dynamics enables the Ising network to settle to a SAT solution if the output spin is set to $+1$.

Three metrics  are important in determining the size/area and routing architecture of an \acs{FPIM}: 1) the number of spins, 2) the number of couplings per spin (sparsity), and 3) the number of bits of coupling resolution (\acs{BCR}) needed.
These metrics depend not only on the specific SAT problem, but also on choices that can be exercised during SAT-to-Ising mapping.
We assess \acs{FPIM} using these metrics on 2000 problems (\texttt{uf20} and \texttt{uf50} sets) from DIMACS SATLIB \cite{Hoos2000SATLIB}, exploring three different choices of technology mapping (gate libraries).
We find that the number of spins needed grows almost linearly\longemdash\/as $\cal O$($n^{1.1}$)\longemdash\/as SAT problem sizes increase. 
Moreover, sparsity  and \acs{BCR} stay at small values that do not grow with the SAT problem's size.
These are significant results, indicating that \acs{FPIM}s can be scaled easily to large \acs{SAT} problem sizes.
Running place and route, we find that all 1000 \texttt{uf20} problems
can be mapped to an \acs{FPIM} with fewer than 6 tracks, and that 9 tracks suffice for the 1000 \texttt{uf50} problems.
Our results suggest that scalable \acs{FPIM}s for SAT will be practical in the near term.

\jnote{Say something about the mapping, \ie, Ising equivalent gates}

\jnote{organization remainder of paper}

\tmpignore{
\jsumm{Introduce the SAT problem, the Ising problem, D-Wave/Chimera, Su/He's work, Ising machines (\ie, classical, esp.\ \acs{OIM})}
\jnote{
\begin{itemize}
\item The SATisfiability problem is a core NP-hard problem with myriad real world applications. Some examples (get from Sharad's review paper).
\item It has been worked on for a long time by many, resulting in heuristic algorithms that mostly do a very good job of solving large real-world problems.
\item Recently, a class of special-purpose analog-hardware-oriented techniques, namely Ising machines, has emerged. Ising machines have generated considerable excitement on account of their promise for solving a wide variety of NP-hard/complete problems more quickly and energy-efficiently than software algorithms or digital hardware.
\item Examples of Ising machines include D-Wave's adiabatic quantum annealer, the Coherent Ising Machine (\acs{CIM})
\item Ising machines represent CO problems in Ising form, which are then solved using said special-purpose hardware.
\item The Ising form essentially consists of a graph, whose vertices, termed spins, take binary values $\pm 1$. The undirected edges of the graph are weighted. The structure of the graph and the values of the edge weights specify the Ising problem to be solved. On this graph is a defined a real-valued, scalar cost function known as the Ising Hamiltonian. Solving the Ising problem consists of finding spin assignments that minimize the Ising Hamiltonian.
\item Since both SAT and Ising are NP problems, any SAT problem can be converted to Ising form \cite{lucas2014ising}.  The Ising representation of a given SAT problem must, however, suit the hardware architecture of the Ising machine, in particular its graph connectivity fabric.
\item Previous work in 2016 \cite{SuTuHeDAC2016SATmappingForDWaveShort} has focussed on the connectivity fabric of D-Wave's quantum annealer, \ie, the Chimera connectivity pattern. 
\item Since then, however, new types of Ising machines that can be fabricated entire on chip in standard CMOS technologies have been developed\longemdash\/in particular, oscillator \cite{OIM} and latch-based \cite{BRIM,BLIM} Ising machines.
\item Such CMOS IC-based Ising machines offer considerably greater flexibility in connectivity, removing the limitation of D-Wave's fixed-connectivity architectures. 
\item Here, we propose FPIM. The main idea: reconfigurable sparse connectivity.
\item Why it matters: minor embedding greatly increases the number of spins needed; but our technique does not.
\end{itemize}
}

\jsumm{What we contribute here: extend the ideas of Su/He to classical Ising machines}
\jnote{Especially those like \acs{OIM} that feature instantaneous dense connectivity on-chip. Advantages: sparsity and low BCR. The concept of FPIM.}

\jsumm{Summary of our results}
}

\ignore{
\begin{wraptable}[24]{r}{0.37\linewidth}
 \setstretch{0.3} 
  \vskip-1em
    \centering
    {\scriptsize
        \begin{tabular}{|p{0.13\textwidth}|p{0.182\textwidth}|}
        \hline
        \raggedright 
        \vfill \acs{BCR}            \vfill & \vfill \mbox{\acl{BCR}}    \vfill    \\ \hline
        \vfill \acs{BM}             \vfill & \vfill \acl{BM}            \vfill    \\ \hline
        \vfill \acs{CIM}            \vfill & \vfill \acl{CIM}           \vfill    \\ \hline
        \vfill \acs{CIMpp}          \vfill & \vfill \acl{CIMpp}         \vfill    \\ \hline
        \vfill \acs{DEaSP}          \vfill & \vfill \acl{DEaSP}         \vfill    \\ \hline
        \vfill \acs{DivConqBCR}     \vfill & \vfill \acl{DivConqBCR}    \vfill    \\ \hline
        \vfill \acs{DivConqSize}    \vfill & \vfill \acl{DivConqSize}   \vfill    \\ \hline
        \vfill \acs{FPIM}           \vfill & \vfill \acl{FPIM}          \vfill    \\ \hline
        \vfill \acs{FPOIM}          \vfill & \vfill \acl{FPOIM}         \vfill    \\ \hline
        \vfill \acs{ImLI}           \vfill & \vfill \acl{ImLI}          \vfill    \\ \hline
        \vfill \acs{MFMCI}          \vfill & \vfill \acl{MFMCI}         \vfill    \\ \hline
        \vfill \acs{MILPI}          \vfill & \vfill \acl{MILPI}         \vfill    \\ \hline
        \vfill \acs{MIMOMLEI}       \vfill & \vfill \acl{MIMOMLEI}      \vfill    \\ \hline
        \vfill \acs{OIM}            \vfill & \vfill \acl{OIM}           \vfill    \\ \hline
        \vfill \acs{OIMLI}          \vfill & \vfill \acl{OIMLI}         \vfill    \\ \hline
        \vfill \acs{QuICCSAT}       \vfill & \vfill \acl{QuICCSAT}      \vfill    \\ \hline
        \vfill \acs{RBM}            \vfill & \vfill \acl{RBM}           \vfill    \\ \hline
        \vfill \acs{xIMSIM}         \vfill & \vfill \acl{xIMSIM}        \vfill    \\ \hline
        \end{tabular}
    }
  \vskip-0.5em
  \caption{\acs{TITLE} acronyms.\label{tab:acronyms}}
\end{wraptable}
}

\ignore{
\begin{wrapfigure}[8]{r}{0.6\linewidth}
    \vskip-2em
    \centering
    \includegraphics[width=\linewidth]{figures/QuICCSAT-flow}
    \vskip-0.0em
    \caption{SAT problem solver stack and flow.\label{fig:SATflow}}
\end{wrapfigure}

\begin{figure}[ht!]
    \centering
    \includegraphics[width=0.85\linewidth]{figures/MIMO-MLE-Ising-flow}
    \vskip-0.0em
    \caption{\acs{MIMOMLE} problem solver stack and flow.\label{fig:MIMOMLEflow}}
\end{figure}
}


\section{Background}
\label{sec:background}

The \acl{SAT} (\acs{SAT}) problem is a fundamental, classically difficult NP-complete/hard CO problem \cite{karp1972np,ZhMa2002questForEfficientSAT} with a wide variety of practical applications \cite{MarquesSilva2002SATpracticalApplications}.
A SAT problem is typically expressed in conjunctive normal form (CNF \cite{ZhMa2002questForEfficientSAT}), where each conjunct is the OR of a set of variables or inverted variables. 
This is a logical expression; the problem is to determine if there exists an assignment of truth values to the variables where the Boolean expression is satisfied.
The Ising model is a general formulation for CO problems \cite{lucas2014ising}, including SAT, based on a cost function
\be{eqn:IsingHamiltonian}
    H(s_1, \cdots, s_n) \triangleq = -\frac{1}{2} \sum_{i=1}^n \sum_{j=1}^n J_{ij} \, s_i s_j,
\ee
called the \textit{Ising Hamiltonian}.
In \cref{eqn:IsingHamiltonian}, $J_{ij}$ is a real-valued weight associated with the connection or coupling between the $i^\text{th}$ and $j^\text{th}$ spins, $n$ is the total number of spins, and $s_i \in \pm 1$ is the value of the $i^{\text{th}}$ spin.\footnote{The weights are symmetric, \ie, $J_{ij} = J_{ji}$; also, $J_{ii} \triangleq 0$. 
Note also that so-called bias or ``magnetic-field'' terms of the form $\sum_i b_i \, s_i$ added to the right-hand side of \cref{eqn:IsingHamiltonian}, are included in \cref{eqn:IsingHamiltonian}, by using an extra spin set to $1$.}
The Ising problem is to find a set of spin assignments that minimizes the Hamiltonian.

In analog Ising machines, spins are implemented using analog circuits.
For example, \ignore{simulated annealing machines \cite{yamaoka2016IsingCMOSshort} implement each spin as a standard binary bit, \eg, using digital registers;} \acl{OIM}s (\acs{OIM}s) \cite{WaWuRoDAC2019OIMlateBreaking} use oscillators whose phases encode spins, BRIM \cite{AfZhVeIgHu2021HPCABRIMshort,RoUCNC2021BLIMcompact} uses ZIV diodes and BLIM \cite{RoUCNC2021BLIMcompact} uses latches, all in analog operation.
Connections between spins, or coupling weights, are typically implemented using resistors.

\ignore{
The most relevant prior work mapping SAT problems to Ising form is that of Su \etal\  \cite{SuTuHeDAC2016SATmappingForDWaveShort}, which targets D-Wave's Chimera spin-connection topology \cite{BoothbyEtAlChimeraPegasusMapping2020Short}.
\redHL{Be more clear about this: our work targets on-chip, ...}
In contrast to D-Wave's fixed connection topologies, however, on-chip CMOS Ising machines offer far greater topological and technological flexibility, as well as mapping efficiency.
}

As noted earlier, to implement an Ising machine on chip at scale, the number of spins in the Ising form of a CO problem should scale only modestly with problem size.
Three further metrics are important: the \textit{number of spins needed}, the \textit{connection sparsity} and the \textit{\acl{BCR} (\acs{BCR})} required to implement the weights.
The number of spins directly determines the number of hardware modules needed on chip.
Sparsity refers to the average number of spins each spin is coupled to.
If a problem (or more precisely, its Ising formulation) is sparse,\footnote{\ie, each spin is coupled to only a few other spins, on average.} then on-chip hardware implementation becomes substantially simpler than if, \eg, the coupling is dense.\footnote{\ie, each spin is coupled to every other spin.}
This is due to the difficulty of routing dense connections, requiring ${n \choose 2} = \frac{n(n-1)}{2}$ wires,\footnote{For example, for $n=1000$ spins, the number of coupling interconnections needed is about half a million.} on chip; this rapidly becomes very difficult as $n$ increases.
However, if the problem is sparse, with average node degree $d \ll n-1$, then the total number of coupling interconnections needed is $n \frac{d}{2}$, which is easily routable.
Indeed, in almost all existing ICs, $d$ is typically a small constant,
much lower than the number of units that need to be connected; for example,
$d$ for a digital CMOS chip is typically between three and four.

\acl{BCR} (\acs{BCR}) refers to the number of bits needed to program the coupling resistors in analog \acs{IM}s.
A BCR of $b$ bits supports $2^b$ different settable coupling values.
The required BCR depends on the nature of the Ising problem to be solved.\ignore{; for example, it has been shown \cite{RoWaSr2021MUMIMOdetectionRSpreprintCompact} that for MU-MIMO wireless detection problems, a minimum \acs{BCR} of about $6$ is needed for good solution quality.}
The complexity of the circuit implementation of each programmable coupling (resistor) depends directly on the \acs{BCR}, since a switched ladder network, requiring $b$ switching elements, is typically employed.
\redHL{Maybe something about the area, too.}

Sparsity and BCR both depend on the problem to be solved and its Ising mapping.
For easy IC implementation, a problem would ideally be sparse and also feature low \acs{BCR}.
\redHL{What are the area implications, both ways? How about quoting specific comparisons from the results section?}
\jnote{UPDATE THIS As we show below, our proposed technique for mapping SAT problems programmably on to IC implementations  offers a range of trade-offs between the two metrics, together with a third one, the total number of spins needed.
As a trend, using more spins leads to better sparsity as well as lower BCR.}

\section{\acs{FPIM}s: \acl{FPIM}s }
\label{sec:FPIMsForSAT}

A central goal of this work is to design an analog Ising chip that can solve problems featuring \textit{different} (sparse) coupling connectivities on the \textit{same} chip.
Prior chip-based approaches towards this goal focussed on mapping problems onto a fixed connectivity topology such as Chimera and Pegasus \cite{BoothbyEtAlChimeraPegasusMapping2020Short} and King's graphs \cite{SugieEtAl2021minorEmbeddingLargeKingsGraphShort,AhChMoKi2021JSSCringOscIMshort}.
\ignore{\rmwords{This is because their target implementation substrate was constrained by the current state-of-the-art in quantum computing hardware. \jnote{Not King's graphs, which Chris Kim used because they couldn't do or even think of anything else. Need to find a way to express this without looking like jerks.}}}
We propose a much more general scheme involving programmable connectivity fabrics, with significant advantages\jnote{  (PROOF? including fewer spins needed for problems of a given size, and lower \acs{BCR})}.
\ignore{\redHL{take out? [Our scheme, which adapts ideas from FPGAs, subsumes the above fixed connectivity topologies as simple special cases; \ie, the same chip will be reconfigurable to different standard connection topologies (including Chimera, Pegasus and King's graphs).]} distracts from the flow, and not important.}
We leverage the flexibility of CMOS, and draw inspiration from a wealth of prior research into the design and implementation of FPGA architectures.
The ideas behind \acs{FPIM}s are developed here in the context of SAT problems, though we anticipate uses for other problem classes as well.

The key steps in our \acs{FPIM} synthesis flow for SAT are 
    1) to synthesize a SAT problem given in CNF (or any other form) as a standard gate level circuit implementation whose output is logic 1 iff the problem is SATisfiable; 
    2) to map this logic circuit implementation into an Ising problem, using ``Ising equivalents'' of the logic gates, with a known ground state\footnote{Minimum-Hamiltonian states are termed ``ground states''.} that provably corresponds to the original problem being SAT; \redHL{See isingmap below} and
    3) to devise an FPGA-like Ising solver chip that connects gates' Ising equivalents reconfigurably.
Sparse connectivity between logic gates translates to sparse connectivity in the Ising equivalent network.
Since widely available logic synthesis tools (\eg, \texttt{yosys}, \texttt{ABC} \cite{wolf2013yosysShort,BrMiCAVABC2010Short})\redHL{ commercial?} are very good at synthesizing sparsely connected, practically routable gate-level circuit implementations of virtually any multi-input logic function, this flow maps \acs{SAT} problems into sparse, routable Ising solver ICs. 
Moreover, the \acs{BCR} needed for the Ising network is proportional to the maximum number of connections to/from any single gate (\ie, maximum fan-in + fan-out) in this scheme.
Since any combinational logic gate/function can be broken down into connections/compositions of \eg, two-input, one-output gates, the typical \acs{BCR} does not increase with the size of the SAT problem; indeed, it can be limited to a low value,  as seen in \cref{sec:results}.
These features of our \acs{FPIM} flow are well suited for practical on-chip implementation at scale.

\subsection{SAT to Ising mapping for flexible on-chip implementation}
\label{sec:SATtoIsing}

\newcommand{\NOT}{!}
\begin{table}[htb]
    \small{
    \centering
    \begin{tabular}{|m{0.21\linewidth}||c|c|c|c|c|c||c|}
        \hline
        {\bf logic}             & $c_{z}$ & $c_{za}$ & $c_{zb}$ & $c_{a}$ & $c_{b}$ & $c_{ab}$ & $H_{\text{SAT}}$ \\ \hline
        \hline
        $z=$ false                & 1       & 0        & 0        & 0       & 0       & 0        & -1               \\ \hline
        $z=\neg a \land \neg b$   & 2       & 2        & 2        & 1       & 1       & 1        & -3               \\ \hline
        $z=a \lor b$              & -2      & -2       & -2       & 1       & 1       & 1        & -3               \\ \hline
        $z=a \land \neg b$        & 2       & -2       & 2        & -1      & 1       & -1       & -3               \\ \hline
        $z=\neg a$                & 0       & 1        & 0        & 0       & 0       & 0        & -1               \\ \hline
        $z=\neg a \lor \neg b$    & -2      & 2        & 2        & -1      & -1      & 1        & -3               \\ \hline
        $z=a \land b$             & 2       & -2       & -2       & -1      & -1      & 1        & -3               \\ \hline
        $z=\neg b \lor a \land b$ & -2      & -2       & 2        & 1       & -1      & -1       & -3               \\ \hline
        $z=a$                     & 0       & -1       & 0        & 0       & 0       & 0        & -1               \\ \hline
    \end{tabular}
    \caption{Ising 2inp technology map. \label{tab:ising2inp}}
    }
\end{table}

We build on ideas developed in \cite{SuTuHeDAC2016SATmappingForDWaveShort,BiChMaRoSeVaArXiv2018SATandMAXSATwQA} for mapping SAT to Ising form.
The first step in our SAT-to-Ising flow is to map a given logical expression for SAT to a gate-level circuit using logic synthesis tools.
Given a user-specified gate library, such tools can realize a Boolean expression as a logic network using gates from the library.
Different problems will, in general, map to gate-level networks with different connectivities and primitive gates from the specified library.

\begin{wrapfigure}[14]{r}{0.45\linewidth}
  \vskip-0.0em
  \centering{
    \scalebox{0.45}{\input{figures/aORb-IsingEqvt-saved-edited.eps_tex}}
  }
 \vskip-1.1em
 \caption{\small Ising equivalent of 2-input OR gate. The spin in the center is fixed at value $+1$.\label{fig:OR2Ising}}
\end{wrapfigure}
Simply realizing a problem's Boolean function as a logic gate-level network does not help solve it for SAT, however.
It is here that conversion to an Ising machine makes the key difference, \ie, the network becomes capable of solving the SAT problem, even though the connection of logic gates it is based on cannot.
To achieve an \acs{IM}, each logic gate maps to a small \textit{Ising equivalent} network, in which the inputs and output of the logic gate are represented by Ising spins. 
The coupling weights between the spins are carefully chosen to correspond to the specific type of logic gate in question (\eg, see \cref{tab:ising2inp}).
\textit{The key property that is met by a correct choice of coupling weights is that if the gate's I/O spins satisfy the Boolean relationship of the desired type of gate, then the Ising Hamiltonian of the Ising equivalent network is the minimum possible, \ie, it is the global minimum.}
The converse is also true, \ie, if spin assignments are not compatible with the logic function of the gate, then the Ising equivalent's Hamiltonian value is \textit{strictly greater} than its minimum. 
Importantly, this property also holds for arbitrary connections of Ising equivalent gates, implying that the overall synthesized Hamiltonian reaches its minimum possible Hamiltonian if, and only if, the logical relationships of all the connected gates are satisfied.

\ignore{
Finally, and crucially, any spin in the coupled Ising equivalent network can be fixed to a desired value. 
Getting the Ising equivalent network to solve SAT then simply translates to fixing its output node to spin value $+1$ ($+1$ represents the logical value \texttt{true}).
}

\jnote{
There is a crucial difference between the Underlying difference between the IM and the logic network: the latter is a unidirectional signal flow (inputs determine outputs, not vice-versa), whereas the IM is inherently bidirectional, with physical properties that lead to Hamiltonian minimization. The operation mechanisms are very different.
}
\jnote{Following \redHL{refer to specific parts of} \cite{SuTuHeDAC2016SATmappingForDWaveShort},\footnote{\redHL{Unlike [1], we do not further map the Ising equivalent gates onto a regular fixed connectivity topology like Chimera or King's, since in an \acs{FPIM}, each needed Ising-equivalent-gate is directly implemented on chip.}} each logic gate in the synthesized circuit has an Ising equivalent (\ie, a small weighted subgraph); connections between logic gates map to their Ising equivalents.
}
\begin{wrapfigure}[8]{R}{0.6\linewidth}
  \vskip-0.9em
  \centering{
    \input figures/x1orx2orx3.tex 
  }
 \caption{\small Example: $z=a+b+c$ synthesized with 2-input OR gates.\label{fig:CNF3var1clauseOR2syth}}
 \vskip0em
\end{wrapfigure}
The crucial step that enables the Ising version of the synthesized gate-level circuit to implement a SAT solver is that its output spin can be set to 1 very easily in hardware\longemdash\/this enforces the SAT condition.
\textit{Fixing the output node to spin 1 constrains the network's minimum Hamiltonian solutions to correspond automatically to SAT solutions; the system's analog dynamics tends to settle naturally to minimum-Hamiltonian solutions}.
No similar solution mechanism is available in the (otherwise closely related) gate-level digital circuit, because it implements a \textit{directional} signal flow graph; ``setting an output'' does not even make conceptual sense.
\jnote{But since ground states of the Ising version (with output fixed to 1) provably correspond to SAT solutions, solving the equivalent Ising problem for a ground state solves SAT. }



\jsumm{isingmap}


We have devised techniques to find Ising equivalents of 1- and 2-input logic gates.\footnote{Owing to DAC's page limits, we are unable to present our gate-to-Ising-equivalent techniques in detail here; a separate publication is planned.}
The general expression for the Hamiltonian of an Ising graph with up to 3 spins is given by
\be{eqn:IsingEqvtHamiltonian}
    H = c_z z + c_{za} z a + c_{zb} z b + c_a a + c_b b + c_{ab} a b,
\ee
where $a$ and $b$ will map to the inputs, and $z$ to the output, of the gate.
Given any 1- or 2-input logic gate (with one output), we find coefficients $c_z$, $c_{za}$, $c_{zb}$, $c_a$, $c_b$ and $c_{ab}$ such that \cref{eqn:IsingEqvtHamiltonian} is strictly minimized \textit{iff} the truth table defining the logic gate is satisfied. \cref{tab:ising2inp} shows the coefficients for the Ising equivalents of a gate library, dubbed \textit{2inp}, that we use in \cref{sec:results}.\footnote{Note that \textit{2inp} includes the standard logic functions AND and OR, as well as 1-input NOT and through-buffer functions.}
The last column of \cref{tab:ising2inp} also shows the minimum Hamiltonian value; by construction, this is achieved \textit{iff} the correct logical relationship between the gate's inputs and output is met.
For the purposes of this paper, it suffices to verify that for each logic gate in \cref{tab:ising2inp}, the coefficients listed indeed correspond to an Ising equivalent, \ie, the minimum Hamiltonian is obtained \textit{iff} the gate's truth table is satisfied.
This check is easily performed by calculating the Hamiltonian \cref{eqn:IsingEqvtHamiltonian} for each entry in each gate's truth table.

\ignore{By comparing each entry of a gate's truth table with the corresponding Hamiltonian value form \cref{eqn:IsingEqvtHamiltonian}, it can be verified that the minimum Hamiltonian is obtained \textit{iff} the truth table is satisfied.}

\begin{figure}[ht]
\vskip-0.0em
    \centering
    \scalebox{0.4}{\input{figures/aORbORc-IsingEqvt-saved-edited.eps_tex}}
    \caption{Ising equivalent of the gate-level circuit in \cref{fig:CNF3var1clauseOR2syth}, with the output spin set to $+1$ to enforce SAT.\label{fig:CNF3var1clauseIsingEqvt}}
    \vskip-0.5em
\end{figure}

We illustrate Ising equivalents and their composition for solving SAT using a simple OR gate example.
\cref{fig:OR2Ising} shows the Ising equivalent of a 2-input OR gate.\footnote{The coefficients in \cref{eqn:IsingEqvtHamiltonian} for OR are the edge weights shown in \cref{fig:OR2Ising}.}
Now consider the following SAT problem in 3 variables $a$, $b$ and $c$, with only one CNF clause:
\be{eqn:CNF3var1clause}
    z \triangleq a + b + c = 1.
\ee
This is easily synthesized using two 2-input OR gates  as
\be{eqn:CNF3var1clauseOR2syth}
    d = a + b, \quad z = c + d.
\ee
as shown in \cref{fig:CNF3var1clauseOR2syth}.
By connecting the OR gates' Ising equivalents in exactly the same manner as in \cref{fig:CNF3var1clauseOR2syth}, and in addition (crucially) setting the output $z$ to spin value $+1$, the Ising-mapped version of the SAT problem \cref{eqn:CNF3var1clause} is obtained, as shown in \cref{fig:CNF3var1clauseIsingEqvt}.\footnote{Multiple edges between spin pairs have been merged (weights summed).}

\subsection{Reconfigurable networks of gate Ising-equivalents: \acs{FPIM}s}
\label{sec:FPIMdetails}

The basic idea of \acs{FPIM} is to implement Ising equivalent networks reconfigurably, using techniques similar to those used in FPGAs.
Instead of LUTs as primitive building blocks (as in FPGAs), \acs{FPIM}s use hardware spin representations (\eg, oscillators for \acs{OIM}, ZIV diodes for \acs{BRIM}, CMOS latches for \acs{BLIM}, \etc) as primitive elements.
In addition to programmable interconnect, \acs{FPIM}s also require programmable weights with a resolution determined by the BCR supported by the \acs{FPIM} architecture.
We use an island-style architecture, where each Ising cluster consists of a small array of spins, each with programmable resistance values between them in a dense sub-array.
For SAT problems, the average node connectivity $d$ can be quite small, as global connectivity between spins is sparse, as seen in \cref{sec:results}.

\begin{figure}[ht]
    \centering
    \includegraphics[width=0.85\linewidth]{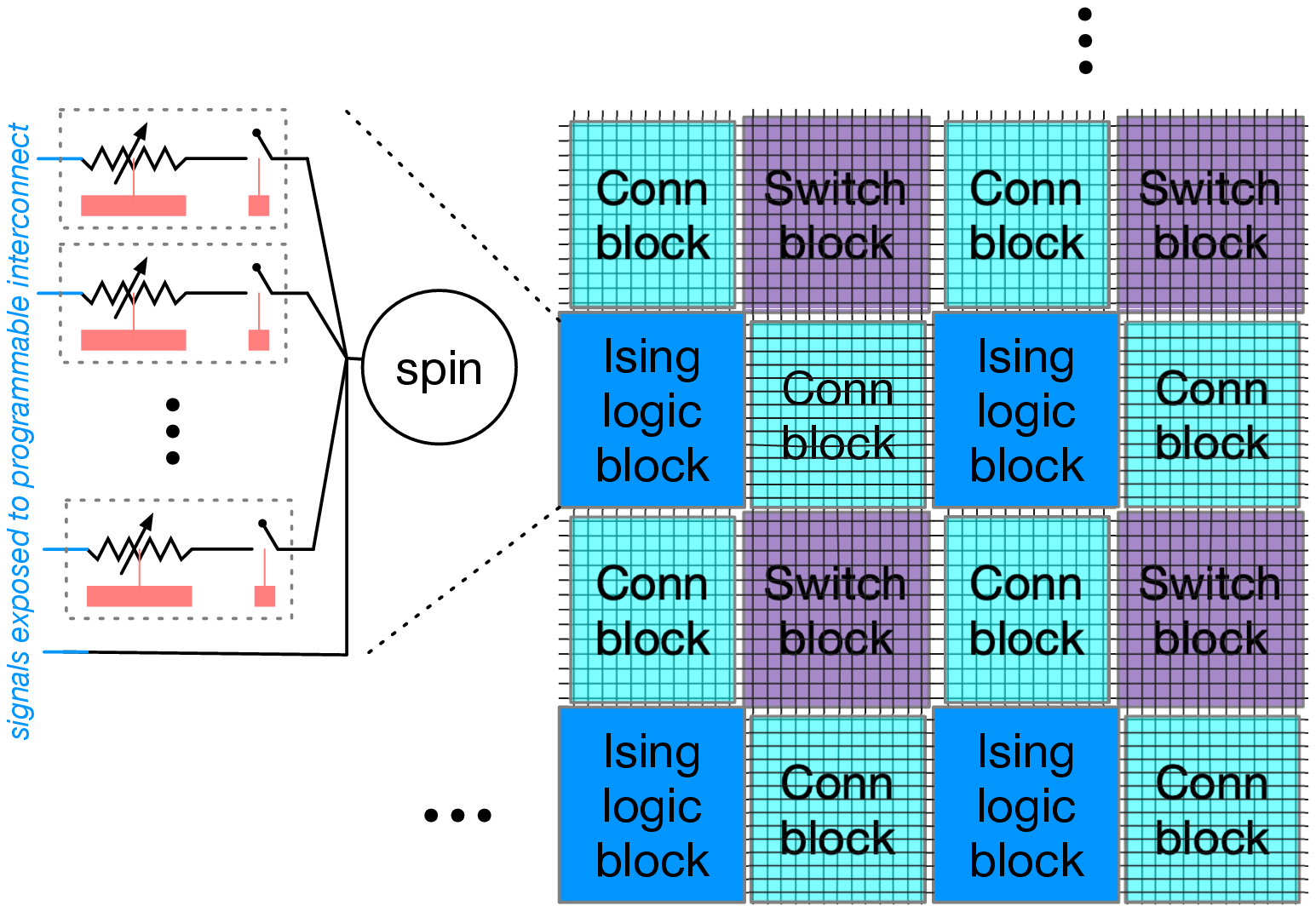}
    \caption{An example island-style \acs{FPIM} architecture. For
      illustration purposes, we have shown one spin per Ising
      block. Programmable state-bits used to configure the
      programmable resistors and coupling switch are shown in
      red. The entire logic block contains a single hardware
      implementation of a spin. A direct connection to the analog
      voltage in the spin is also a primary bidirectional pin for the
      logic block. We use an oscillator for our OIM-based
      FPIM in our evaluation.\label{fig:FPIM}}
\vskip-1em
\end{figure}
There are at least four points of departure for \acs{FPIM}s compared to traditional \acs{FPGA} architectures.
First, our results (see \cref{sec:results}) show a slightly higher connectivity requirement compared to traditional \acs{FPGA}s, which can be compensated for by increasing the number of routing tracks.
Second, connectivity requirements for coupling resistors are bi-directional, which requires pass-transistor based configuration rather than the direct drive architectures seen in modern \acs{FPGA}s.
Third, much larger pass transistors will be needed to ensure that coupling resistance values are primarily determined by programmable resistors rather than the interconnect.
Fourth, a suitable choice of logic block I/O connectivity must be selected for the local island in the island-style architecture.

\ignore{
\redHL{bit of a flow disconnect here - try pushing this to later} 
\rmtolookat{FPGAs, which feature programmable connectivity fabrics that connect logic gates or LUTs,\footnote{Look-up Tables, which can be programmed to mimic different gates.} are well suited for implementing such gate-level circuits.
The architecture of FPGAs has been optimized over almost four decades to be suitable for implementing the output of logic synthesis tools.
}
}

An example overall architecture with one spin per logic block is shown in \cref{fig:FPIM}.
Note that this is a logical diagram, not a physical one because the wires will be partially routed over the Ising logic block.
The connection box and switch box will contain programmable cross-points implemented with large pass transistors to reduce the resistance of the programmable interconnect relative to the explicit coupling resistors required.
Switch boxes permit programmable connectivity between horizontal and vertical routing tracks, while connection boxes provide connectivity between the Ising logic block and the interconnect tracks.
We can use pass-transistors to build switches that have negligible additive resistance values compared to the desired coupling values.
We also observe that the bi-directional nature of coupling in \acs{FPIM} implies that the \acs{FPGA} architecture and interconnect problem for \acs{FPIM} is not exactly the same as that in traditional synchronous \acs{FPGA}s, where the switch boxes can be used to split and fanout digital signals to multiple end-points without any loss/impact on correct operation.
However, the routing problem is similar to the point-to-point pipelined interconnect required in high-throughput asynchronous \acs{FPGA}s \cite{TeMa2004FPGApipelinedAsyncFPGAs}.

Finally, if a spin has extremely high connectivity, we can introduce {\it mirror spins\/} and force these spins to match each other.
This permits decomposition of the small number of spins that have large connectivity for embedding into the \acs{FPIM} architecture.

In a traditional FPGA architecture, the programmable routing fabric dominates the area, delay, and energy of the FPGA.
While we will show that the area overhead of routing is slightly lower than a traditional FPGA for FPIMs, the energy and delay metrics for FPIMs are quite different. 
In particular, the goal of the programmable routing is to resistively couple {\it slow, time-varying, analog voltages\/} to each other, rather than transmitting a digital signal transition. 
The programmable resistors that are part of the logic block are designed to have significantly higher resistance values (20$\times$ or more) than the pass transistor logic for the interconnect, permitting us to treat the interconnect resistance as negligible. 
Hence, the performance and power consumption of the entire system will be dominated by the non-linear dynamics of the spin-coupled analog Ising hardware\longemdash\ie, an FPIM will be logic-block dominated, unlike FPGAs that are interconnect dominated. 
The key question to answer is: can we design an interconnect that is routable, \ie, supports a wide range of Ising-mapped SAT problems? We focus on this in the evaluation next.

\section{\acs{FPIM} metrics on \acs{SAT} benchmarks}
\label{sec:results}

We mapped two sets of SAT problems, \texttt{uf20} and \texttt{uf50} from DIMACS SATLIB \cite{Hoos2000SATLIB} (each consisting of 1000 problems), into Ising form using the above techniques.
\texttt{uf20} problems each have 20 variables and 91 clauses, while \texttt{uf50} problems feature 50 variables and 218 clauses.
Employing \texttt{yosys}/\texttt{ABC}, we synthesized each problem using three different technology libraries: 1) the \textit{2inp} set of logic gates shown in \cref{tab:ising2inp}, 2) NAND gates only, and 3) OR and NOT gates only.\footnote{Being logically complete, NAND and OR-NOT are capable of synthesizing any SAT problem.}
Note that all the Ising equivalent weights for gates in the \textit{2inp} library are between -2 and +2, which bodes well for the \acs{BCR} of our Ising formulation of SAT.

The Ising equivalent network is generated from the SATLIB problem (in \texttt{.cnf} format) via automated custom tools we have written. 
These tools perform gate-level synthesis using \texttt{yosys} and \texttt{ABC}, calculate Ising equivalents for the gates needed (\eg, like those in \cref{tab:ising2inp}) and use them to translate the gate-level netlist into an Ising-equivalent one.
\keepRedHL{details? cnf2rtl.pl (.cnf to .va); run\_yosys.pl (.va to gate netlist .va); to\_act.pl (.va to .act); THEN WHAT?}

To assess hardware requirements for \acs{FPIM}s, we study the characteristics of the Ising formulation of SAT problems that directly impact hardware resource requirements for a programmable architecture.
In particular, we looked at the distributions of the number of spins needed, the sparsity of connections (degree of each spin node), and the \acs{BCR} needed to program each weight value. \redHL{BCR is a tricky one to explain properly}
The distributions over the 1000 problems in each set are shown in \crefrange{tab:uf20-50-spins}{tab:uf20-50-BCRs}.
Each table shows distributions for the three technology mappings used.

It can be seen in \cref{tab:uf20-50-spins} that over the 1000 problems in the \texttt{uf20} benchmark set, the majority (more than 900) require between 151 and 200 spins if the \textit{2inp} technology library is applied.\footnote{The numbers in parentheses in \cref{tab:uf20-50-spins} are for a modified mapping, explained below.}
However, if the NAND-only and OR-NOT-only libraries are used, the range shifts to 251-300 and 301-350, respectively.
The sparsities (average number of nodes each node is connected) over all the \texttt{uf20} problems are shown in \cref{tab:uf20-50-degrees}.
The table reveals that all three technology mappings have roughly the same sparsity characteristics, with the majority of nodes having between 1 and 5 connections.
\ignore{\rmwords{Note that constant spin values for Ising problems are similar to constant values in Boolean logic, and don't require high fanout support as they can be locally generated as necessary. {\bf IS THIS TRUE?} \jnote{probably not completely - I don't think global routing of ``clock'' spins can be avoided.}}
\redHL{say something about the fixed 1 node requiring a sort of global clock routing? Does it make sense to divide the node count by the total number of spins in each problem?}
\greenHL{interesting, maybe?}}
The \acs{BCR} distribution, \ie, the count of connections that need a given number of bits to program their weight values, is shown in \cref{tab:uf20-50-BCRs}.
\ignore{\redHL{again, normalizing by the total number of spins or connections per problem might provide more insight. Increment BCR counts by 1! (for the sign bit). Clarify w Rajit what exactly 0 means, anyway.} \greenHL{0 means it is 1 or -1. So if we didn't have a sign bit, this would be always 1\longemdash\ie, zero bits.}}
It can be seen that 3 bits of coupling resolution suffice for the majority of connections over all the problems, with a smaller number of 4 and 5 bit connections also needed; and that these requirements are largely independent of the technology mapping used.
This constitutes important information for planning and designing an \acs{FPIM} for \texttt{uf20} problems.

\newcommand\B[1]{{\color{blue}#1}}
\newcommand\G[1]{{\color{mydarkgreen}#1}}
\newcommand\R[1]{{\color{MyDarkRed}#1}}

\begin{table}[htb]
    \small{
    \centering
    \setlength{\tabcolsep}{0.38em} 
    \begin{tabular}{|m{0.15\linewidth}||c|c|c||c|c|c|}
        \hline
        \hfill lib$\rightarrow$ & {\footnotesize\B{2inp}} & {\scriptsize\G{NAND}}  & {\scriptsize\R{N/OR}} & {\footnotesize\B{2inp}} & {\scriptsize\G{NAND}} & {\scriptsize\R{N/OR}} \\ \hline
        {\bf \#\ spins}         & \multicolumn{3}{c||}{\texttt{uf20}} & \multicolumn{3}{c|}{\texttt{uf50}} \\ \hline
        \hline
        0-150     & \B{1 (0)}    &          &          &         &         &                        \\ \hline
        151-200   & \B{760 (346)}  &          &          &         &         &                        \\ \hline
        201-250   & \B{239 (654)} & \G{61}   &          &         &         &                        \\ \hline
        251-300   &  \B{0 (29)}       & \G{845}  & \R{76}   &         &         &                        \\ \hline
        301-350   &          & \G{94}   & \R{862}  &         &         &                        \\ \hline
        351-400   &          &          & \R{62}   &         &         &                        \\ \hline\hline
        500-600   &          &          &          & \B{922 (557)}   &         &                        \\ \hline
        601-701   &          &          &          & \B{78 (443)} &         &                        \\ \hline
        701-801   &          &          &          &       & \G{27}  &                        \\ \hline
        801-901   &          &          &          &         & \G{783} & \R{411}                \\ \hline
        901-1000  &          &          &          &         & \G{190} & \R{589}                \\ \hline
    \end{tabular}
    \caption{Problem counts \vs\ number of Ising-mapped spins. The
      numbers in parentheses for the 2inp library correspond to
      values after spin mirroring. N/OR = OR-NOT library.\label{tab:uf20-50-spins}}
    }
    \vspace{-1em}
\end{table}

Similar data on the 1000 \texttt{uf50} problems is shown in the latter columns of \cref{tab:uf20-50-spins,tab:uf20-50-degrees,tab:uf20-50-BCRs}.
Since these have more Boolean variables than \texttt{uf20}, it is not surprising that larger numbers of spins are required in their Ising mappings.
Like for \texttt{uf20}, the \textit{2inp} library results in significantly more compact mappings from a number-of-spins perspective, with every problem mapped using between 601 and 800 spins.
We determined empirically that the number of Ising spins grows as $\cal O$($n^{1.1}$) with respect to the number of SAT variables $n$.
Note, however, that the sparsity (\cref{tab:uf20-50-degrees}) and \acs{BCR} (\cref{tab:uf20-50-BCRs}) distributions are very similar to those of \texttt{uf20}.
This provides the important insight that \textit{while the number of spins increases as SAT problem sizes increase, the number of connections each spin requires, as well as the number of bits required to program those connections, remains roughly the same.}
This directly implies that scalable \acs{FPIM} architectures for arbitrary SAT-to-Ising problem sizes are possible without having to scale up connectivity and \acs{BCR} requirements with problem size.

\begin{table}[htb]
    \small{
    \centering
    \setlength{\tabcolsep}{0.45em} 
    \begin{tabular}{|m{0.15\linewidth}||c|c|c||c|c|c|}
        \hline
        \hfill lib$\rightarrow$ & {\footnotesize\B{2inp}} & {\scriptsize\G{NAND}}  & {\scriptsize\R{NOT/OR}} & {\footnotesize\B{2inp}} & {\scriptsize\G{NAND}} & {\scriptsize\R{NOT/OR}} \\ \hline
        {\bf degree}            & \multicolumn{3}{c||}{\texttt{uf20}} & \multicolumn{3}{c|}{\texttt{uf50}} \\ \hline
        \hline
        1-5       & \B{160.1}  & \G{225.9} & \R{225.9}   &  \B{501.4} & \G{749.6}  &  \R{783.7}     \\ \hline
        6-10      & \B{12.6}   & \G{36.4}  & \R{36.7}    &  \B{22.3}  & \G{60.6}   &  \R{60.8}      \\ \hline
        11-20     & \B{16.3}   & \G{14.1}  & \R{14.2}    &  \B{23.3}  & \G{58.4}   &  \R{58.5}      \\ \hline
        21-31     & \B{2.5}    & \G{0.01}  & \R{0.013}   &  \B{25.3}  & \G{3.489}  &  \R{1.124}     \\ \hline
        32-41     &            &           &             &  \B{0.817} &            &                                  \\ \hline
    \end{tabular}
    \caption{Average number of nodes/problem \vs\ node degree. \label{tab:uf20-50-degrees}}
    }
    \vskip-1em
\end{table}

\begin{table}[htb]
    \small{
    \centering
    \setlength{\tabcolsep}{0.45em} 
    \begin{tabular}{|m{0.11\linewidth}||c|c|c||c|c|c|}
        \hline
        \hfill lib$\rightarrow$  & {\footnotesize\B{2inp}} & {\scriptsize\G{NAND}}  & {\scriptsize\R{NOT/OR}} & {\footnotesize\B{2inp}} & {\scriptsize\G{NAND}} & {\scriptsize\R{NOT/OR}} \\ \hline
        {\bf BCR}                 & \multicolumn{3}{c||}{\texttt{uf20}} & \multicolumn{3}{c|}{\texttt{uf50}} \\ \hline
        \hline
        1   & \B{223.4}  & \G{319.3} & \R{476.4} &      \B{745.3}  & \G{1056}    &  \R{1375.0}  \\ \hline
        2   & \B{357.0}  & \G{409.0} & \R{457.9} &      \B{1070.5} & \G{1300.6}  &  \R{1326.9}  \\ \hline
        3   & \B{110.6}  & \G{133.0} & \R{20.0}  &      \B{287.01} & \G{318.3}   &  \R{32.645}  \\ \hline
        4   & \B{0.798}  & \G{14.2}  & \R{14.1}  &      \B{4.301}  & \G{56.015}  &  \R{56.074}  \\ \hline
        5   & \B{0.001}  & \G{0.085} & \R{0.087} &      \B{0.06}   & \G{3.489}   &  \R{3.535}   \\ \hline
    \end{tabular}
    \caption{Avg.\ number of weights/problem \vs\ \acs{BCR} needed. \label{tab:uf20-50-BCRs}}
    }
    \vskip-1em
\end{table}

\jnote{
for each of 3 technologies:
What does the BCR graph show? number of weights over all problems for each absolute value of coupling? what are the decimal points - divided by the total number of problems = 1000?
What was BCR=0?
Usefulness: tells you how much hardware you need to plan: so many connections that are 2-bit programmable, so-many 3-bit, etc.
What about just bunches hardwired coupling resistors with different relative coupling values?
They could be chosen as part of the connectivity fabric? (``coarse coupling choice'')
Plus, a few programmable connections with some BCR (``fine coupling choice''). 
What is a BCR of 0 (or, w sign bit, 1)?
}

\ignore{
\rmtolookat{\redHL{this probably goes away}
We analyzed the connectivity for challenging 50-variable SAT problems using our SAT mapping flow \redHL{below}; it is extremely sparse.
\cref{fig:degreedistSAT} shows the degree distribution of a typical 50-variable challenging SAT problem from the DIMACS SATLIB benchmark suite.
In a SAT problem that requires 883 spins, 761 (86\%) of these spins have five or fewer coupling resistors with the average number of resistors per spin being 5.33.
The optimized version of the same SAT problem requires 581 spins, and 516 (89\%) of them have five or fewer coupling resistors with the average number of resistors per spin being 6.46.
\redHL{As we show in the results section, }this connectivity requirement is only mildly larger than the connectivity requirements in traditional FPGAs.
Hence, programmable interconnect architectures based on conventional \acs{FPGA}s are an excellent starting point for designing \acs{FPIM}s.
}

\begin{figure}[ht]
    \vskip-1em
    \centering
    \includegraphics[width=0.8\linewidth]{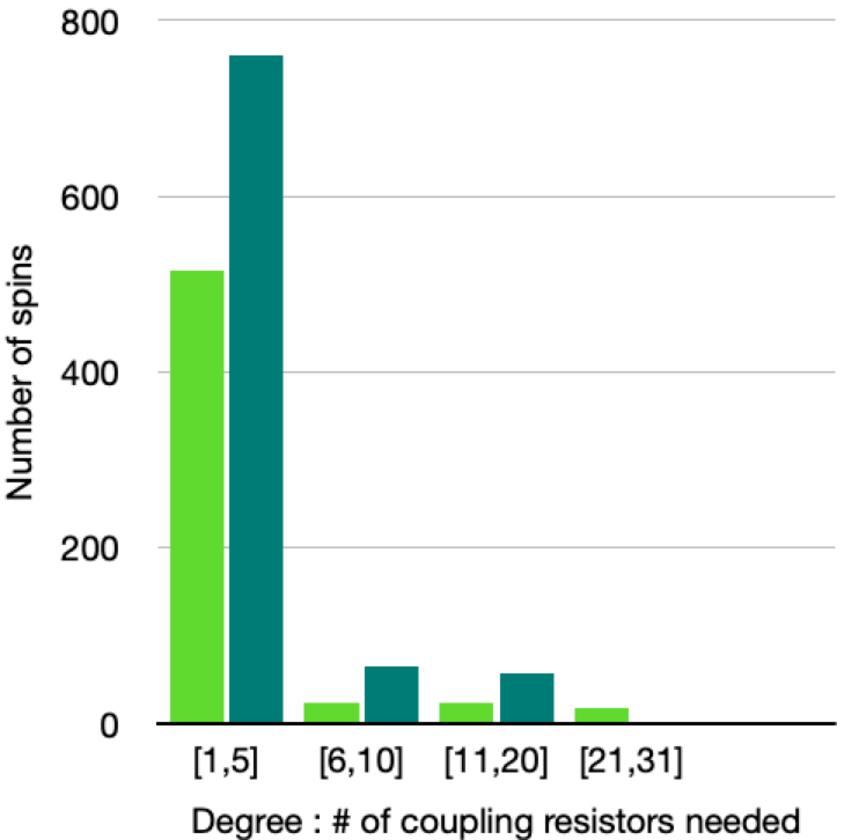}
    \caption{\redHL{probably to be removed} A typical distribution of connectivity between spins for challenging 50-variable SAT problems for both the unoptimized (dark green) and optimized (light green) Ising library. The vast majority of spins have five or fewer coupling resistors.\label{fig:degreedistSAT}}
\end{figure}
}

\ignore{ replaced by tables
\begin{figure}[ht]
    \vskip-1em
    \centering
    \includegraphics[width=0.99\linewidth]{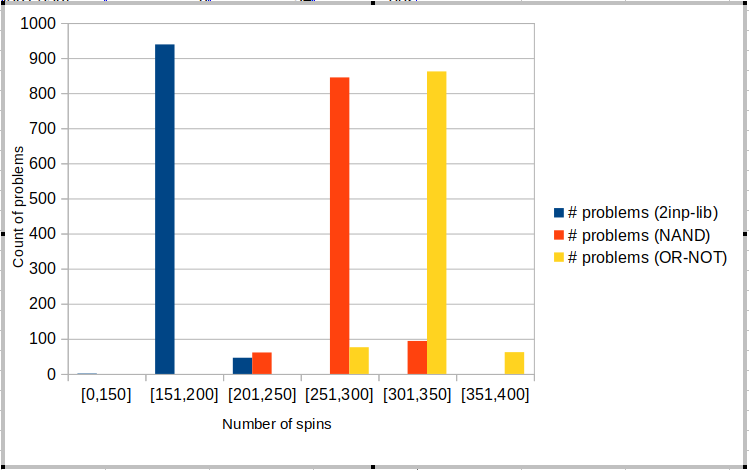}
    \caption{\keepRedHL{uf20-numspins}.\label{fig:uf20numspins}}
\end{figure}

\begin{figure}[ht]
    \vskip-1em
    \centering
    \includegraphics[width=0.99\linewidth]{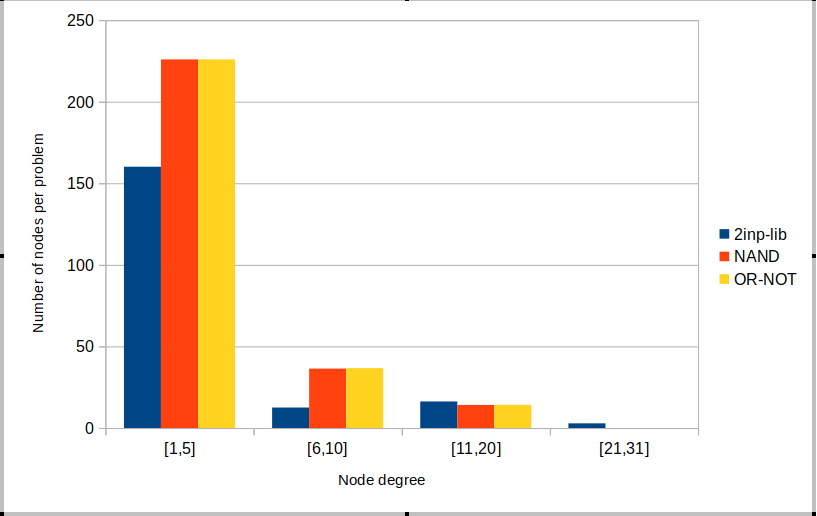}
    \caption{\keepRedHL{uf20-node-degrees}.\label{fig:uf20degrees}}
\end{figure}

\begin{figure}[ht]
    \vskip-1em
    \centering
    \includegraphics[width=0.99\linewidth]{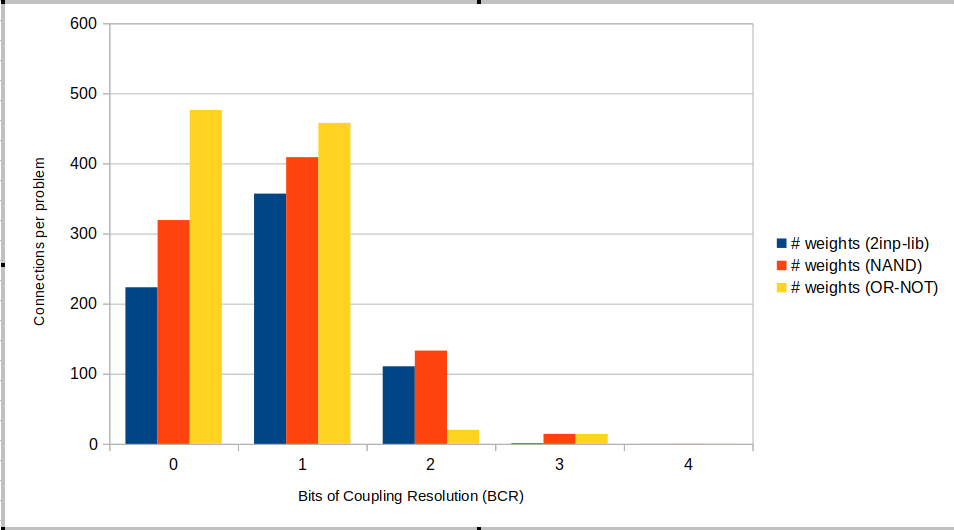}
    \caption{\keepRedHL{uf20-BCRs}.\label{fig:uf20BCRs}}
\end{figure}

\begin{figure}[ht]
    \vskip-1em
    \centering
    \includegraphics[width=0.99\linewidth]{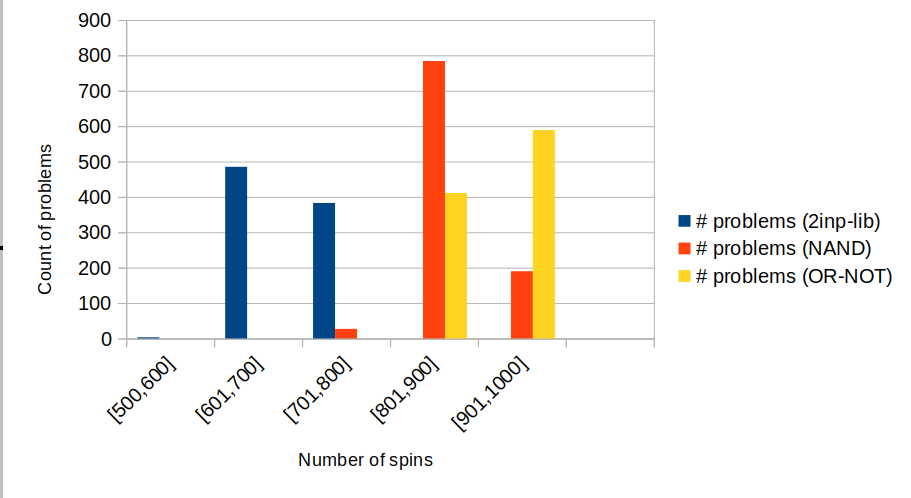}
    \caption{\keepRedHL{uf50-numspins}.\label{fig:uf50numspins}}
\end{figure}

\begin{figure}[ht]
    \vskip-1em
    \centering
    \includegraphics[width=0.99\linewidth]{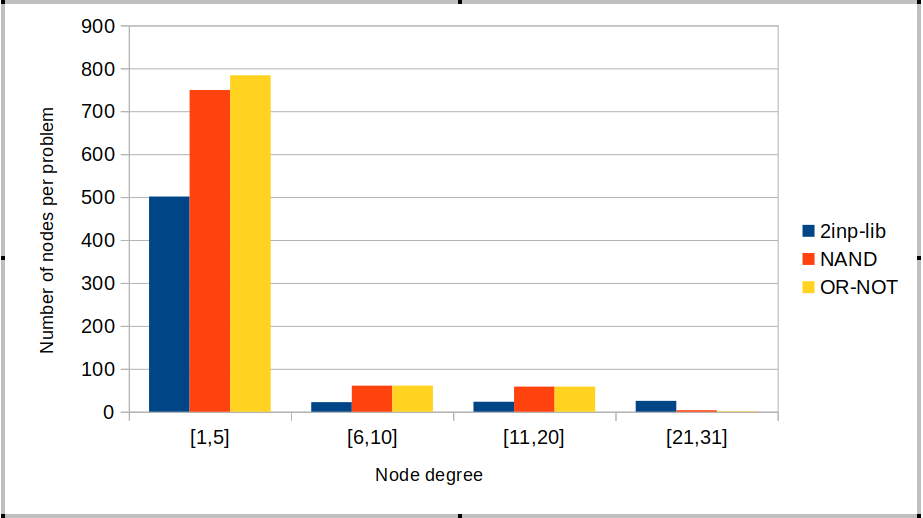}
    \caption{\keepRedHL{uf50-node-degrees}.\label{fig:uf50degrees}}
\end{figure}

\begin{figure}[ht]
    \vskip-1em
    \centering
    \includegraphics[width=0.99\linewidth]{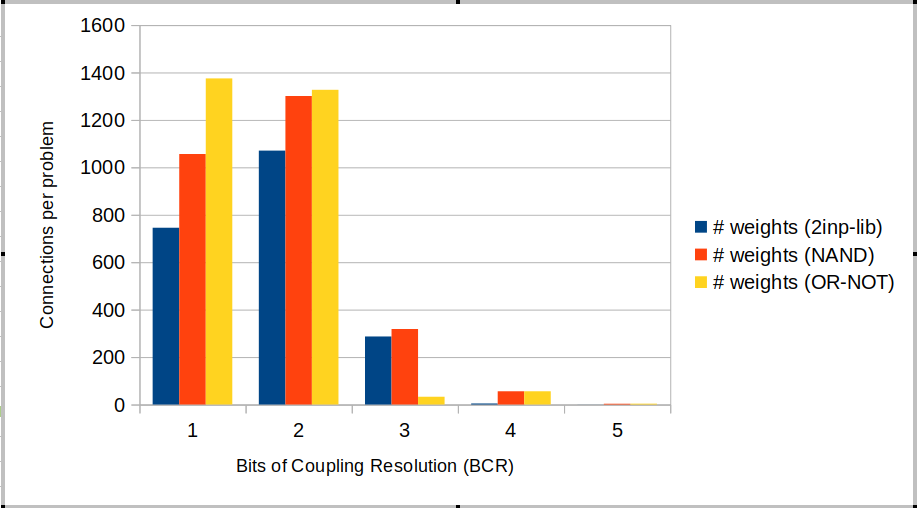}
    \caption{\keepRedHL{uf50-BCRs}.\label{fig:uf50BCRs}}
\end{figure}
}

\begin{figure}[ht]
    \vskip-0em
    \centering
    \includegraphics[width=0.85\linewidth]{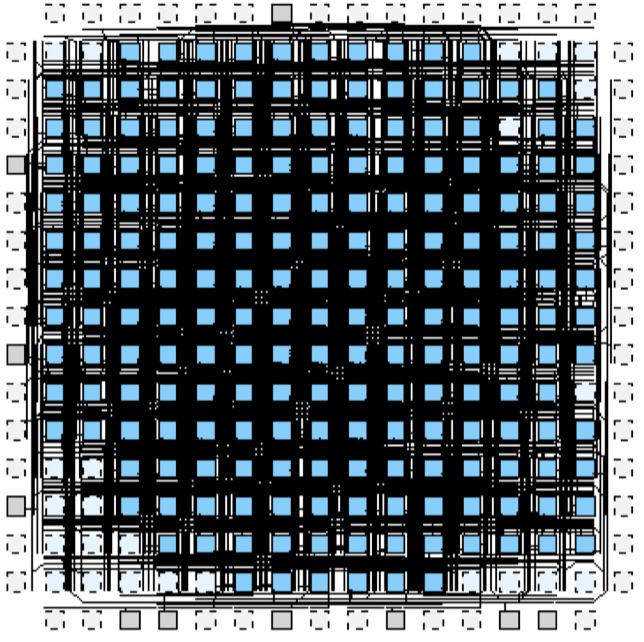}
    \caption{Placement/routing of example \texttt{uf20} problem.\label{fig:vpruf2001}}
\end{figure}

To determine the interconnect complexity required for an \acs{FPIM}, we used the open-source VPR FPGA place-and-route software package \cite{BeRo1997vprShort}.
We created an FPGA architecture description file for VPR that treats each spin as a place-able component. 
Connectivity is determined by the Ising problem generated via our SAT-to-Ising automated mapping flow.  
The interconnect has half-populated connection blocks, with a standard Wilton-style switch block, with direct connections to resistors connected to spins. 
The routing tracks are all singles (\ie, single-hop per switch point). 
We ran the \texttt{uf20} and \texttt{uf50} problems through place-and-route to determine the number of routing tracks necessary to support each problem.
\cref{fig:vpruf2001} shows an example \texttt{uf20} problem placed and routed on an \acs{FPIM}\ignore{; \cref{fig:vpruf20closeup} shows a more detailed view of a section of the architecture}.
\ignore{\newRedHL{needs updates}} All the \texttt{uf20} benchmarks can be mapped to an \acs{FPIM} with less than 20 routing tracks, and all the \texttt{uf50} benchmarks can be mapped to an \acs{FPIM} with less than 36 routing tracks.
While these numbers are slightly \ignore{\newRedHL{NOT?}} higher than
conventional FPGAs given the problem size, they are well within the capabilities of a modern CMOS process.

\begin{figure}[ht]
    \vskip-0em
    \centering
    \includegraphics[width=0.85\linewidth,height=1.5in]{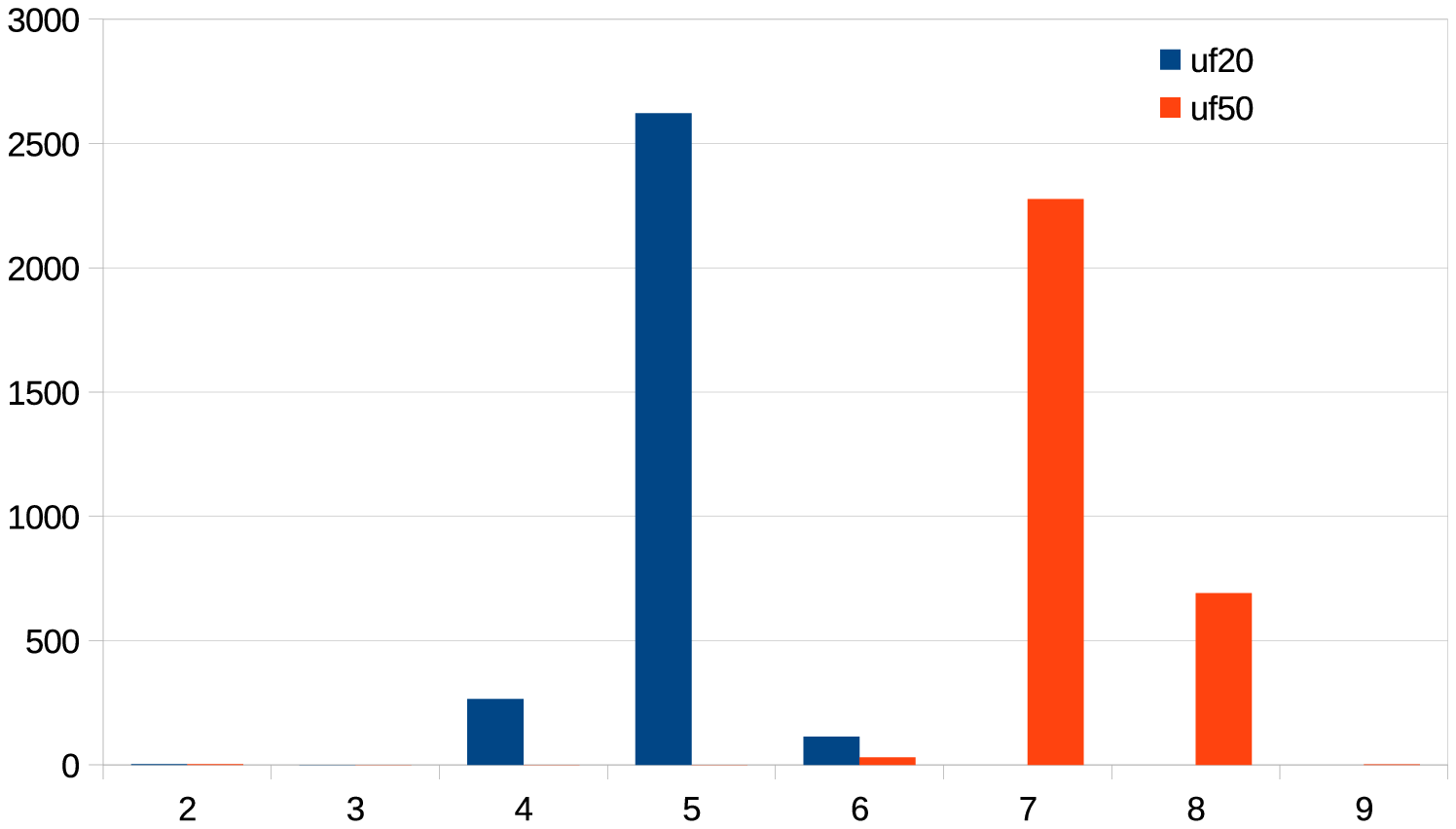}
    \caption{\texttt{uf20} and \texttt{uf50}: routing tracks needed by problems.\label{fig:track-hist}}
    \vskip-0.5em
\end{figure}

To reduce the interconnect requirements at the cost of slightly
increased spins, we incorporated a spin mirror step after the Ising
problems were created.  The spin mirror step examines spins that have
high connectivity (more than 8), and introduces a replica spin to
reduce the connectivity in the Ising formulation where the replica is
generated by inserting a buffer gate prior to Ising
mapping. \ignore{\newRedHL{similar to minor embedding?}}  The numbers
in parentheses in \cref{tab:uf20-50-spins} show the increase in spins
as a result of this transformation.  This is similar to the process of
buffer insertion in digital logic to reduce delay, but our motivation
is to reduce the complexity of the programmable interconnect.  We
re-ran VPR on the 6,000 benchmarks (2,000 problems, 3 libraries per
problem).  We found that all the \texttt{uf20} benchmarks were
routable with 6 tracks for a 6.3\% increase in number of spins, and
all the \texttt{uf50} benchmarks were routable with 9 tracks for a
3.9\% increase in number spins\longemdash\/a 3.3$\times$-4$\times$
reduction in the number of routing tracks.  \cref{fig:track-hist}
shows the histogram of routing track requirements across all libraries
and benchmarks (3,000 scenarios for \texttt{uf20} and for
\texttt{uf50}).

Using the above information, we have estimated the areas of all the components needed to make an \acs{FPIM}, including analog spin circuits,\footnote{assuming ring oscillators, which require more devices than, \eg, CMOS latches.} programmable resistors implemented using transmission gate ladders, and transmission gate based switch boxes, connection boxes and input MUXes.
The overall estimate came to about 10M$\lambda^2$ per spin, where $\lambda$ represents half of the technology's feature size.
Unlike in FPGAs, this area is strongly dominated by the analog components, \ie, the analog spins and programmable coupling resistors.
In a 65nm process, for example, a 1000-spin \acs{FPIM}, which is more than adequate for all the \texttt{uf50} and \texttt{uf20} problems, would occupy about 10mm$^2$.
This shows that \acs{FPIM}s are in fact a practical way to implement configurable IM solvers.

\ignore{
\begin{figure}[ht]
    \vskip-0em
    \centering
    \includegraphics[width=0.99\linewidth]{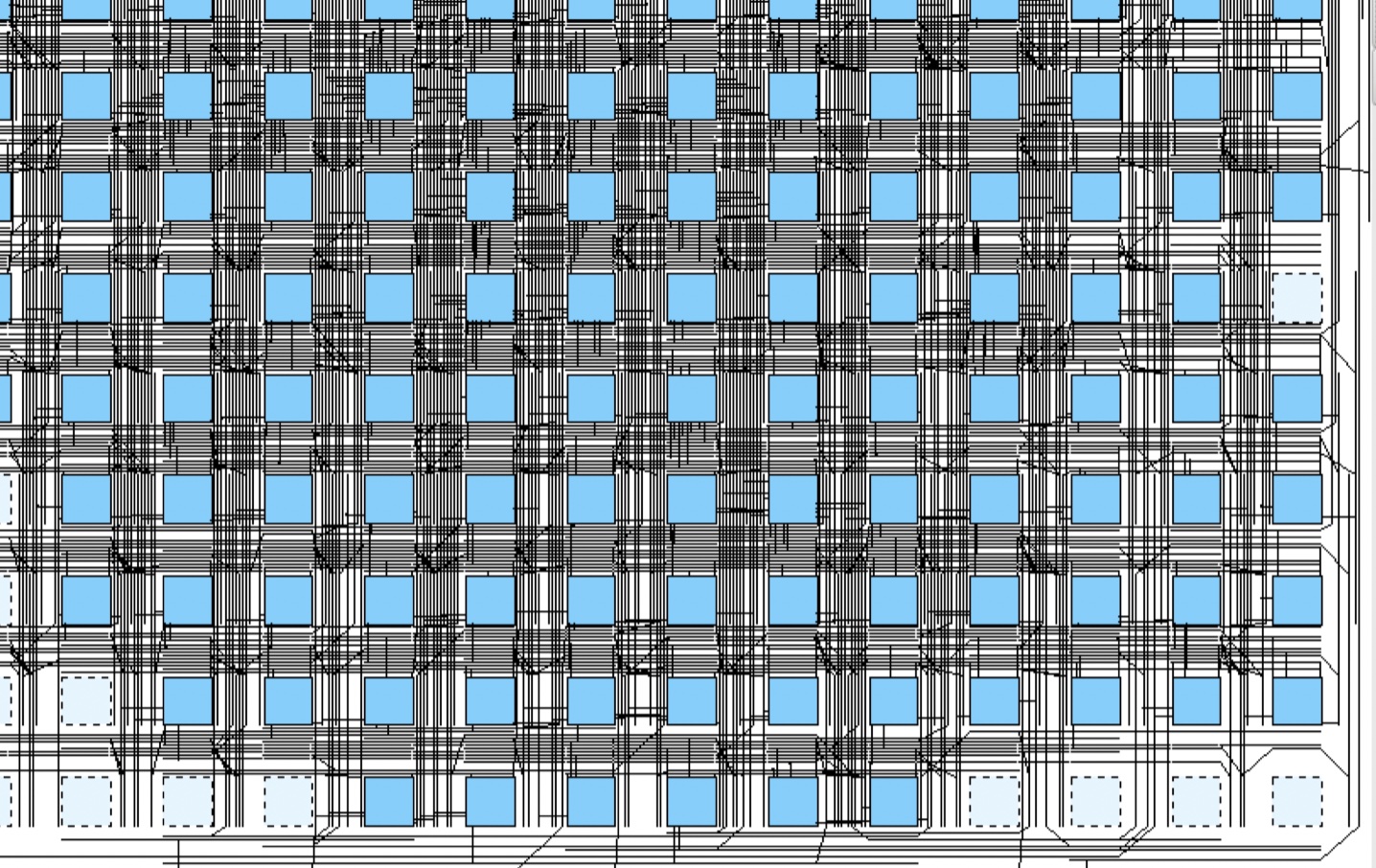}
    \caption{Placement/routing of example \texttt{uf20} problem (detail).\label{fig:vpruf20closeup}}
\end{figure}
}

\subsection*{Conclusion}
\jnote{area evolving to the realization that one size does not fit all and that practical IC Ising solver architectures must be tailored to the problem to be practically effective. }
Specializing analog Ising solver ICs\ignore{has been evolving to the realization that one size does not fit all and that practical IC Ising solver architectures} for problem classes is likely a more practically feasible and competitive approach than re-mapping onto fixed, regular Ising connection topologies.
We have shown that SAT problems can be mapped onto a novel but very realistic reconfigurable analog Ising solver architecture, \acs{FPIM}.
We have assessed feasibility using key metrics: the number of spins needed, connection sparsity, and the number of bits needed to program connection weights (BCR).
\ignore{
\newRedHL{say something about these metrics!}
We have shown that the \texttt{uf20} and \texttt{uf50} SATLIB benchmark sets can map to \acs{FPIM}s with 6 and 9 routing tracks respectively, easily feasible in today's technologies. \redHL{summary of area/blocks (Rajit) - in 65nm technology}
}
\acs{FPIM} is not limited to a specific electronic spin circuit, but can employ spin circuits from several available analog Ising schemes.
A 1000-spin FPIM, which suffices for all 2000 SAT benchmark problems we considered, would occupy about $10$mm$^2$ in a 65nm process.
\acs{FPIM} is a significant step towards practical analog Ising solver ICs for SAT and possibly other CO problems.

\let\OLDthebibliography\thebibliography
\renewcommand\thebibliography[1]{
  \OLDthebibliography{#1}
  \setlength{\parskip}{0pt}
  \setlength{\itemsep}{2pt plus 0.3ex}
}

{
\bibliographystyle{IEEEtran}
\setlength\columnsep{0.5em}
\renewcommand{\baselinestretch}{0.8}
\footnotesize
\bibliography{stringdefs,jr,PHLOGON-jr,tianshi,verification-hybrid-systems-CPS}
}

\end{document}


\title[Subtitle, if you have one]
    {LaTeX2e guide for authors using the \cambridge\ design}

  \author{ALI WOOLLATT\\[3\baselineskip]
    This guide was compiled using \hbox{\cambridge.cls \version}\\[\baselineskip]
    The latest version can be downloaded from:
    https://authornet.cambridge.org/information/productionguide/
      LaTeX\_files/\cambridge.zip}

  \frontmatter
  \maketitle
  \tableofcontents
  \listoffigures
  \listoftables
  \listoffloatingboxes
  \listofcontributors

\chapter*{Notation}

\begin{tabular}{@{}ll}
AFM & atomic force microscope\\
AKPZ & anisotropic KPZ equation\\
$a_0$ & lattice constant\\
$c_q(\ell)$ & q-th order correlation function\\
$d_{\mathrm {E}}$ & embedding dimension\\
$d_f$ & fractal dimension\\
$L$ & system size \\
$\equiv$ & {\em defined} to be equal\\
$\sim$ & {\em asymptotically} equal (in scaling sense)\\
$\approx$ & {\em approximately} equal (in numberical value)
\end{tabular}

  \mainmatter
  \part{Getting started}

\chapter{Introduction}
\label{intro}

This guide is for authors who are preparing a book for Cambridge University Press using the \LaTeX\ document preparation system, and the \cambridge\ class file.

The \LaTeX\ document preparation system is a special version of the \TeX\ typesetting program. \LaTeX\ adds to \TeX\ a collection of commands which simplify typesetting by allowing the author to concentrate on the logical structure of the document rather than its visual layout.

\LaTeX\ provides a consistent and comprehensive document preparation interface. There are simple-to-use commands for generating a table of contents (toc), lists of figures and/or tables, and indexes. \LaTeX\ can automatically number list entries, equations, figures, tables, and footnotes, as well as parts, chapters, sections and subsections. Using this numbering system, bibliographic citations, page references and cross references to any other numbered entity (e.g. chapter, section, equation, figure, list entry) are quite straightforward.

\LaTeX\ is a powerful tool for managing long and complex documents. In particular, partial processing enables long documents to be produced chapter by chapter without losing sequential information. The use of document classes allows a simple change of style to transform the appearance of your document.

\section{The \LaTeXe\ book document class}

The \cambridge\ class file preserves the standard \LaTeX\ interface such that any document which can be produced using the standard \LaTeXe\ book class can also be produced with the \cambridge\ class. However, the measure (i.e. width of text) is different from that for book, therefore linebreaks will change and long equations may need re-setting.

\section{The \cambridge\ document class}

The \cambridge\ design has been implemented as a \LaTeXe\ class file, and is based on the book class as discussed in the \LaTeX\ manual. Commands which differ from the standard \LaTeX\ interface, or which are provided in addition to the standard interface, are explained in this guide. This guide is \emph{not} a substitute for the \LaTeX\ manual itself.

\section{Implementing the \cambridge\ class file}
\label{usingcamb}

Copy \cambridge.cls into the correct subdirectory on your system. The \cambridge\ document class is implemented as a complete document class, \emph{not} a document class option. To run this guide through \LaTeX, you need to include the following class and style files:\\[0.5\baselineskip]
\verb"  \documentclass{"\texttt{\cambridge}\verb"}"\\
\verb"    \usepackage[rightcaption,raggedright]{sidecap}% for side captions"\\
\verb"    \usepackage{framed} % for floatingboxes"\\
\verb"    \usepackage{soul}   % for letterspacing in theorem-style headings"\\
\verb"    \usepackage[agsm]{harvard}"\\
\verb"    \usepackage{rotating}"\\
\verb"    \usepackage{floatpag}"\\
\verb"      \rotfloatpagestyle{empty}"\\
\verb"    \usepackage{amsthm}"\\
\verb"    \usepackage{graphicx}"\\
\verb"    \usepackage{multind}\ProvidesPackage{multind}"\\[0.5\baselineskip]
It may be that your book does not use references, rotation, theorems, graphics, or multiple indexes, in which case you simply need the first line. If you include \verb"multind.sty", you must also insert the command \verb"\ProvidesPackage{multind}". More recent style files include this information; it simply sends a message to the class file to re-style the index into the \cambridge\ style.

In general, the following standard document class options should \emph{not} be used:
 \begin{itemize}
  \item \texttt{10pt}, \texttt{11pt}, \texttt{12pt};
  \item \texttt{oneside}  (\texttt{twoside} is the default);
  \item \texttt{fleqn}, \texttt{leqno}, \texttt{titlepage}, \texttt{twocolumn}.
 \end{itemize}

\section{Implementing the multi-contributor option}

This option should be used where chapters have been written by different contributors. Please read Section~\ref{usingcamb} first; then implement the \verb"[multi]" option as follows:\\[0.5\baselineskip]
\verb"  \documentclass[multi]{"\texttt{\cambridge}\verb"}"\\[0.5\baselineskip]
Further details can be found in Section~\ref{multicontributor}.

\section{Fonts}

The typefaces for the final typeset version of the \cambridge\ design are Times for the text, and Adobe Helvetica Neue Condensed for the sans-serif elements, such as headings.

It is a good idea to start working with these fonts straight away; you will get an idea of the final look of the text, and you will know the extent. If you cannot use the Adobe font for any reason, it is acceptable to default to the standard Times sans-serif.

If your book is going to be typeset by Cambridge University Press, you are welcome to submit your files using Computer Modern; we will change the font. Authors supplying final PDFs must use Times.

\subsection{Times}
We recommend you use one of the following versions of Times:
\begin{enumerate}
\item mathptmx, available from:\\
      http://www.ctan.org/tex-archive/fonts/psfonts/psnfss-source/mathptmx/
\item txfonts, available from:\\
      http://www.ctan.org/tex-archive/fonts/txfonts/
\end{enumerate}
Mathptmx changes the default roman font to Adobe Times, but does not support bold math characters.

Txfonts does support bold math, but the kerning of subscripts and superscripts is not ideal. You must load txfonts \emph{after} amsthm.sty, otherwise you will get some `already defined' messages.\footnote{The reason we do not include times.sty as an option is because it mixes Computer Modern and Times fonts, and there is a clash between math and italic characters.}

\subsection{Adobe Helvetica Neue Condensed}

This typeface is available to purchase in OpenType format from Adobe. If you have this typeface and are able to convert it to a \LaTeX-usable format, include it by adding the \verb"[prodtf]" option as follows:
\begin{verbatim}
  \documentclass[prodtf]{EngC}
\end{verbatim}
This command will call in helvneue.sty, which is distributed with this package.

\section{Submission of files}
Please note that you must supply a PDF of your files so that the typesetters
can check characters such as bold math italic. If you are providing final PDF files
for printing, remember to embed all fonts as Type~1 fonts.

\section{Make-up}

This is a generic guide for many Cambridge designs. We have therefore not attempted to correct long lines, and there are occasions where pages may be a little long. The latter is due to the use of \verb"\begin{samepage}"\ldots \verb"\end{samepage}" where we are keeping text together for clarity. Authors should not include any page make-up commands, unless they are providing final PDFs for printing.




  \alphafootnotes
  \author[M\,M Magn\'usson and D\,A Tranah]
    {Magn\'us M\'ar Magn\'usson\footnotemark\
    and David Tranah\footnotemark\\
    International Glaciological Society}

  \chapterauthor{Magn\'us M\'ar Magn\'usson\footnotemark\
    and David Tranah\footnotemark
    \affil{International Glaciological Society}}

  \chapter{The \cambridge\ class file in detail}

  \footnotetext[1]{Formerly of the Icelandic
    Meteorological Office, Reykjav\'\i k.}
  \footnotetext[2]{Supported by NSF Grant 43645.}
  \arabicfootnotes

  \contributor{Magn\'us M\'ar Magn\'usson
    \affiliation{International Glaciological Society,
      Scott Polar Research Institute,
      Lensfield Road, Cambridge CB2 1ER}}

  \contributor{David Tranah
    \affiliation{Cambridge University Press,
      The Edinburgh Building, Shaftesbury Road,
      Cambridge CB2 8RU}}


The following notes may help you achieve the best effects with the \cambridge\ class file.

\section{Frenchspacing}

The \verb"\frenchspacing" option has been selected by default. This ensures that no extra space is inserted after full points, and is normal practice. If there is a strong reason for reversing this, you can key \verb"\nonfrenchspacing" in the preamble.

\section{Adding a subtitle to the front page}

The standard \verb"\title" command has been extended to take an optional argument which is then used as a subtitle on the main title page. For example, this document uses following title command:
\begin{verbatim}
  \title[Subtitle, if you have one]
    {LaTeX2e guide for authors using the \cambridge\ design}
\end{verbatim}

\section{Adding a blank page to your document}

Blank pages should not be numbered. If you require one, use the command \verb"\cleardoublepage", which has been redefined to start the next page on a recto, and if necessary, insert a totally blank verso page first.

\section{Chapter numbering}
If your book starts with an unnumbered chapter (e.g. \verb"\chapter*{Introduction}", then make all the numbered elements (e.g. section heads) unnumbered, by using \verb"\section*{...}". Otherwise, sections will be numbered 0.1, 0.2, etc.

\section{Section numbering}

\LaTeX\ provides five levels of section heads, and they are all defined in the \cambridge\ class file: \verb"\section", \verb"\subsection", \verb"\subsubsection", \verb"\paragraph", and \verb"\subparagraph". Numbers are given for the first three headings.

You can reduce the level of numbered section heads (it is not advisable to increase them). For instance, if you only want headings numbered down to subsections, add the following line to the preamble: \verb"\setcounter{secnumdepth}{2}". To number down to sections, make this \verb"\setcounter{secnumdepth}{1}", etc.

\section{Specifying running heads and toc entries}

\subsection{Single-contributor books}
\label{singlecontributor}

In \cambridge, chapter titles and section heads are used as running heads at the top of every page:
\begin{itemize}
\item chapter titles appear on even-numbered pages (versos), and
\item section heads appear on odd-numbered pages (rectos).
\end{itemize}
A problem with the standard version of \LaTeX\ has always been that the shortened versions of chapter and section titles, specified for running heads, have also been the entries for the toc. There are packages such as the memoir class which enable you to specify different toc entries, running head entries, and chapter titles. However, there is a simple way to add the verbose version of the chapter or section heads into the toc:
\begin{verbatim}
  \chapter[Toc entry]{Verbose chapter title}
  \chaptermark{Running head entry}

  \section[Toc entry]{Verbose section title
    \sectionmark{Running head entry}}
    \sectionmark{Running head entry}
\end{verbatim}
Note that for sections, you need the optional argument to \verb"\section", even if `Toc entry' is in fact the same text as `Verbose section title'. Also, the \verb"\sectionmark" has to be entered twice as shown, because the first \verb"\sectionmark" deals with the header of the page that the \verb"\section" command falls on, and the second deals with subsequent pages.

\subsection{Multi-contributor books}
\label{multicontributor}

Using the \cambridge\ multi-contributor option, author(s) name(s) and chapter titles are used as running heads at the top of every page:
\begin{itemize}
\item author(s) name(s) appear on even-numbered pages (versos), and
\item chapter titles appear on odd-numbered pages (rectos).
\end{itemize}
The author(s) names(s) may run to several lines, and contain new line commands (e.g. \verb"\\"), but the running head must be a single line. To enable you to specify an alternative short form of the author(s) name(s), the standard \verb"\author" command has been extended to take an optional argument to be used as the running head:
\begin{verbatim}
  \author[Author(s) name(s)]{The full author(s) name(s)}
\end{verbatim}
The following shows some coding for a chapter written by two authors, each of whom have footnotes. In this example, the authors' names will immediately follow the chapter title, and will read Magn\'us M\'ar Magn\'usson$^{a}$ and David Tranah$^{b}$. Their respective footnotes will be `$^{a}\enskip$Formerly of the Icelandic Meteorological Office, Reykjav\'\i k.' and `$^{b}\enskip$Supported by NSF~Grant 43645.' It is crucial that \verb"\author" precedes \verb"\chapter". If the authors have footnotes, you must start the chapter with \verb"\alphafootnotes", fill in the details for author(s), chapter title and author footnotes, then key \verb"\arabicfootnotes" to revert to arabic footnotes:
\begin{verbatim}
  \alphafootnotes
  \author[M\,M Magn\'usson and D\,A Tranah]
    {Magn\'us M\'ar Magn\'usson\footnotemark\
    and David Tranah\footnotemark\\
    International Glaciological Society}

  \chapter[Running head entry]
    {The \cambridge\ class file in detail}

  \footnotetext[1]{Formerly of the Icelandic
    Meteorological Office, Reykjav\'\i k.}
  \footnotetext[2]{Supported by NSF Grant 43645.}
  \arabicfootnotes
\end{verbatim}
Note that for multi-contributor books, the long version of the chapter title will always appear in the table of contents.

\section{Adding author(s) name(s) in single-contributor books}
Sometimes, chapters in single-contributor books are written by different people. If you wish the authors (and their affiliations) to appear beneath the chapter opening, as demonstrated in this chapter, key your chapter head as follows; note that \verb"\chapterauthor" must precede \verb"\chapter":
\begin{verbatim}
  \alphafootnotes
  \chapterauthor{Magn\'us M\'ar Magn\'usson\footnotemark\
    and David Tranah\footnotemark
    \affil{International Glaciological Society}}

  \chapter{The \cambridge\ class file in detail}

  \footnotetext[1]{Formerly of the Icelandic
    Meteorological Office, Reykjav\'\i k.}
  \footnotetext[2]{Supported by NSF Grant 43645.}
  \arabicfootnotes
\end{verbatim}
If you have footnotes associated with the authors, you will also need to insert \verb"\alphafootnotes" and \verb"\arabicfootnotes".

\section{List of contributors}
\label{contrib}
The code for generating an automatic list of contributors should be entered after the author and chapter titles, as follows:
\begin{verbatim}
  \contributor{Magn\'us M\'ar Magn\'usson
    \affiliation{International Glaciological Society,
      Scott Polar Research Institute,
      Lensfield Road, Cambridge CB2 1ER}}

  \contributor{David Tranah
    \affiliation{Cambridge University Press,
      The Edinburgh Building, Shaftesbury Road,
      Cambridge CB2 8RU}}
\end{verbatim}
You then simply need to add the \verb"\listofcontributors" command after the table of contents (or after the artwork lists, if included) in the preamble, as follows:
\begin{verbatim}
  \tableofcontents
  \listoffigures
  \listoftables
  \listofcontributors
\end{verbatim}

\subsection{Note to editors regarding the list of contributors}

The contributors will appear in the same order as they are called in, since the list is generated in the same way as the table of contents. This means that at the final stage, the file will require editing to make the entries alphabetic.

Once you have a complete list of contributors, comment out the line which is generating them, and replace it as shown below:
\begin{verbatim}
  \tableofcontents
  \listoffigures
  \listoftables
 %\listofcontributors
  \editedlistofcontributors
\end{verbatim}
Next, rename the file with the extension \verb".loc" to \verb"editedloc.tex" (in the case of this guide, you would rename \texttt{\cambridge guide.loc} to \verb"editedloc.tex"). Edit this file as required, then run the file through \LaTeX\ once more, and the edited version will appear.

\section{Adding an abstract}
The following code will give you an unnumbered section head `Abstract'. Keep the abstract to one paragraph:
\begin{verbatim}
  \begin{abstract}
    Thermal convection driven by centrifugal...
  \end{abstract}
\end{verbatim}

\section{Adding an extract}
You may add an extract -- the following coding:
\begin{verbatim}
  \begin{extract}
    In a semiparametric model, some aspects of the data
    distribution are specified in terms of a small number of
    parameters, but other aspects are left arbitrary.
  \end{extract}
\end{verbatim}
will produce:
  \begin{extract}
    In a semiparametric model, some aspects of the data
    distribution are specified in terms of a small number of
    parameters, but other aspects are left arbitrary.
  \end{extract}

\section{Adding a `copyright' line to a chapter opening~page}
If you are publishing a single chapter, with permission from Cambridge University Press, you may be required to add a copyright line (and/or other information) to the footer of the chapter opening page. This may be achieved using:
\begin{verbatim}
  \copyrightline{Reprinted from \textit{Mathematical
    Methods for Physics and Engineering} by Riley,
    Hobson and Bence \copyright\ 2009 Cambridge
    University Press.}
\end{verbatim}
Should the following chapter not require the copyright line, reverse this immediately before the next \verb"\chapter" command by using:
\begin{verbatim}
  \copyrightline{}
\end{verbatim}

\section{Changing the level of entries in the table of~contents}
\label{changingentries}
The \cambridge\ design will, by default, list parts, chapters and sections in the table of contents. However, to improve the usefulness of this guide, we have used the command:
\begin{verbatim}
  \setcounter{tocdepth}{2}
\end{verbatim}
to increase this by one level, so the table of contents in this document also shows subsections.

\section{Lists}
\label{lists}

The \cambridge\ class provides the following standard list environments:
\begin{enumerate}
 \item numbered lists, created using the \verb"enumerate" environment;
 \item bulleted lists, created using the \verb"itemize" environment;
 \item labelled lists, created using the \verb"description" environment.
\end{enumerate}
The \verb"enumerate" environment numbers each list item with an arabic numeral followed by a full point; alternative styles can be achieved by inserting a redefinition of the number labelling command after the \verb"\begin{enumerate}". For example, a list numbered with lower-case letters inside parentheses can be produced. Because `(a)' is wider than a standard arabic digit, the label width has to be increased. This is achieved by specifying the widest label in the list inside square braces:
\begin{verbatim}
  \begin{enumerate}[(a)]
    \renewcommand{\theenumi}{(\alph{enumi})}
    \item estimate the fluctuations in the near-wall region\ldots
    \item subdue these near-wall fluctuations\ldots
  \end{enumerate}
\end{verbatim}
This produces the following list:
  \begin{enumerate}[(a)]
    \renewcommand{\theenumi}{(\alph{enumi})}
    \item estimate the fluctuations in the near-wall region\ldots
    \item subdue these near-wall fluctuations\ldots
  \end{enumerate}

\section{Endnotes}

In addition to footnotes,\footnote{The footnote counter will be reset on chapters.} the \cambridge\ class provides a similar facility for endnotes. Their appearance depends on which option you are using:
\begin{enumerate}
\item for single-contributor books, the endnotes will be produced in the form of an unnumbered chapter at the end of the book;
\item for multi-contributor books, they are an unnumbered section at the end of each chapter.
\end{enumerate}
Endnotes are inserted into the text in a similar way to footnotes, but using the \verb"\endnote" command; for example,
\begin{verbatim}
  When the Richardson number\endnote{Lewis Fry Richardson
  (1881--1953).\label{richardson}} increases\ldots
\end{verbatim}
will produce `When the Richardson number\endnote{Lewis Fry Richardson (1881--1953).\label{richardson}} increases\ldots' in the text. Authors must choose between using footnotes and endnotes; do not use both.

\subsection{Single-contributor books}
Endnotes should be printed at the end of the book, after the appendices but before the bibliography and/or references.
\begin{verbatim}
    :
  \theendnotes
  \begin{thebibliography}{99}
    :
\end{verbatim}
The \verb"\theendnotes" command generates an unnumbered chapter which appears in the table of contents (see page~\pageref{richardson} for style).

\subsection{Multi-contributor books}

Endnotes should be printed at the end of the chapter using the same \verb"\theendnotes" command.

\section{Examples}
\label{examples}
Examples have rules both above and below. If you require two or more to appear together, simply add another \verb"\item" within the \verb"examplelist" environment (e.g. Example~\ref{fourier}):
\begin{verbatim}
  \begin{examplelist}
    \item Show that the geometrical definition of grad leads to the
          usual expression for $\nabla\phi$ in Cartesian coordinates.
    \solution
          Consider a small rectangular volume element $\Delta V =
          \Delta x\,\Delta y\,\Delta z$ with its faces parallel to the
          $x,y,z$ coordinate surfaces and with the point $P$ at one
          corner. We must calculate\ldots

    \item Find the Fourier series of $f(x) =  x^3$ for $0 < x \leq 2$.
    \label{fourier}
    \solution
          In the example discussed in the previous section we found
          the Fourier series for $f(x) = x^2$ in the required range.
          So, if we \textit{integrate} this term by term\ldots
  \end{examplelist}
\end{verbatim}
This will produce:
  \begin{examplelist}
    \item Show that the geometrical definition of grad leads to the
          usual expression for $\nabla\phi$ in Cartesian coordinates.
    \solution
          Consider a small rectangular volume element $\Delta V =
          \Delta x\,\Delta y\,\Delta z$ with its faces parallel to the
          $x,y,z$ coordinate surfaces and with the point $P$ at one
          corner. We must calculate\ldots

    \item Find the Fourier series of $f(x) =  x^3$ for $0 < x \leq 2$.
    \label{fourier}
    \solution
          In the example discussed in the previous section we found
          the Fourier series for $f(x) = x^2$ in the required range.
          So, if we \textit{integrate} this term by term\ldots
  \end{examplelist}

\section{Problems}

Authors may use the \verb"problemlist" environment which will typeset problems at the end of each section or chapter. This is shown in the following example:
\begin{verbatim}
  \begin{problemlist}
    \item Show that the link between shock formation and
          film rupture is invoked here because of the\ldots
    \item Show that the physical interpretation of\ldots
          \label{physi}
  \end{problemlist}
\end{verbatim}
which will produce:
  \begin{problemlist}
    \item Show that the link between shock formation and
          film rupture is invoked here because of the\ldots
    \item Show that the physical interpretation of\ldots
          \label{physi}
  \end{problemlist}
As with all numbered environments, individual problems (e.g. Problem~\ref{physi}) may be cross-referenced.

\section{Boxes}

Boxes may be typeset using the following coding. They are effectively floats, and so can take the optional arguments [h], [t], [b], as used for figures and tables. The following coding produces Box~\ref{fb}:
\begin{verbatim}
  \begin{floatingbox}[h]
    \caption{Floating box title}
      There are grounds for cautious optimism that we may now
      be near the end of the search for the ultimate laws of
      nature -- Stephen Hawking
    \label{fb}
  \end{floatingbox}
\end{verbatim}
  \begin{floatingbox}[h]
    \caption{Floating box title}
      There are grounds for cautious optimism that we may now
      be near the end of the search for the ultimate laws of
      nature -- Stephen Hawking
    \label{fb}
  \end{floatingbox}

\section{Figures}

The \cambridge\ class file will cope with most positioning of your figures. As captions normally fall below figures, the figure must be included first, then the caption, then the label. This is illustrated in Figure~\ref{cantor}. The \verb"cantor1.eps" file has been called in by using \verb"\usepackage{graphicx}" in the preamble. Note that if you are producing a list of illustrations (using \verb"\listoffigures"), you need to repeat the caption in square braces, but without the full point.

\subsection{Figures between two-thirds and full text width}

\begin{figure}
    \includegraphics[width=\textwidth]{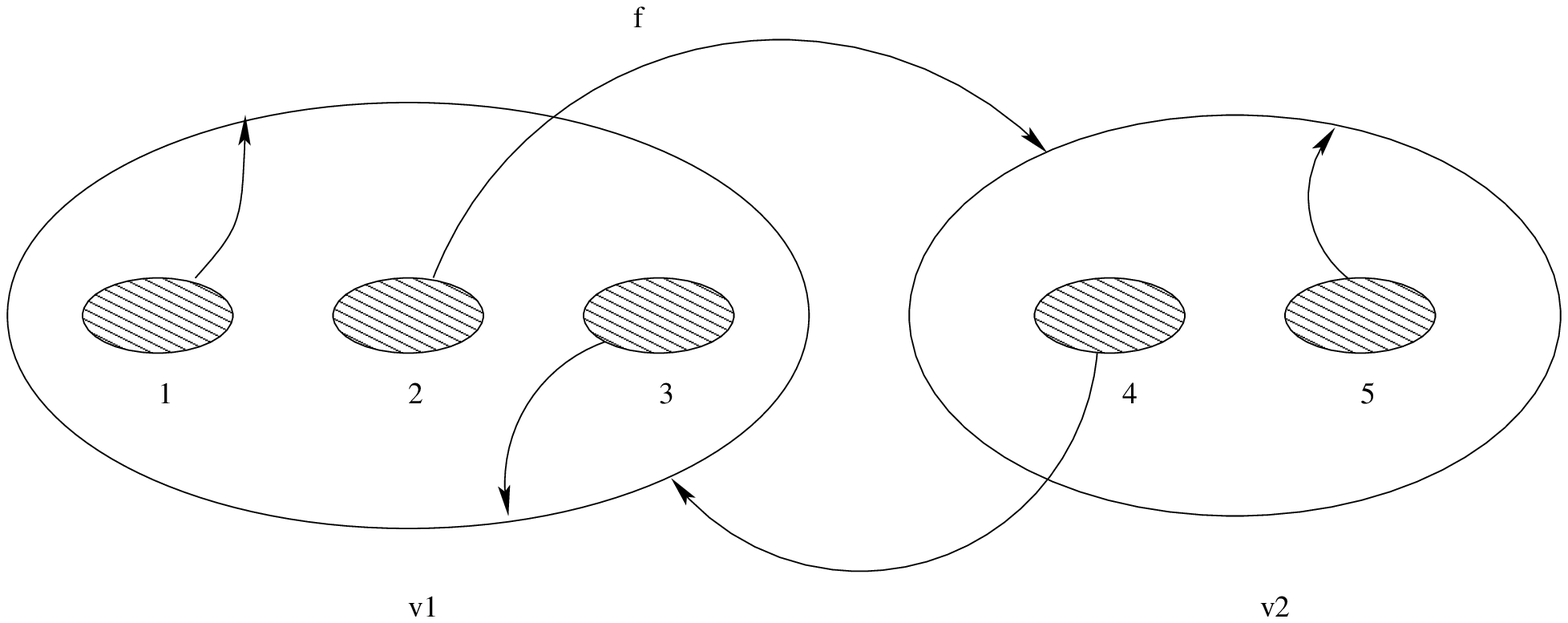}
    \caption[Shortened figure caption for the list of illustrations]
      {A Cantor repeller. Figure captions will be left-aligned,
      29pc wide, and unjustified.}
    \label{cantor}
\rule[-20pt]{\textwidth}{0.5pt}
\begin{verbatim}
  \begin{figure}
    \includegraphics[width=\textwidth]{cantor1.eps}
    %  note that the square brace option below is only required
    %  if you intend to produce a list of illustrations
    \caption[Shortened figure caption for the list of illustrations]
      {A Cantor repeller. Figure captions will be left-aligned,
      29pc wide, and unjustified.}
    \label{cantor}
  \end{figure}
\end{verbatim}
\rule[20pt]{\textwidth}{0.5pt}
  \end{figure}

If the width of the figure lies between two-thirds and the full text width (in other words, between 19pc and 29pc), the figure may be typeset using standard \LaTeX\ coding (see Figure~\ref{cantor}).

\subsection{Figures less than two-thirds of the text width}

If you have a figure which takes up less than two-thirds of the text width, i.e. less than 19pc, you have two options. You may either have the caption below the figure, or choose the space-saving option of placing it to the side.

\subsubsection{Caption below figure}

In this case, you would use the \verb"figure" environment (see Figure~\ref{cantor}) and size your figure accordingly.

\subsubsection{Caption to the side of figure}
\label{captiontoside}

\begin{SCfigure}[50]
    \includegraphics[width=18pc]{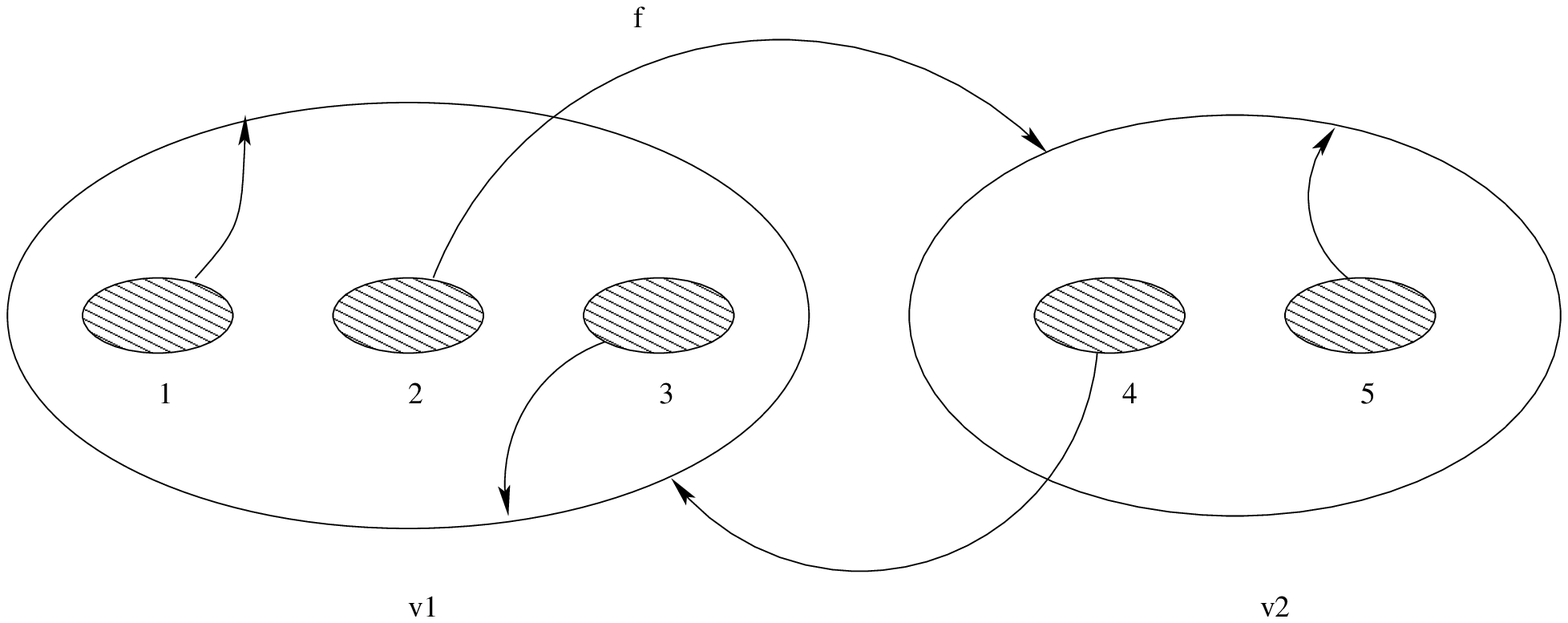}\\[17pt]%
    \hbox to0pt{\rule{\textwidth}{0.5pt}}\\%
    \hbox to0pt{\verb"  \begin{SCfigure}[50]"}\\[-2pt]%
    \hbox to0pt{\verb"    \includegraphics[width=18pc]{cantor1.eps}"}\\[-2pt]%
    \hbox to0pt{\verb"      \caption[Figure with side caption]"}\\[-2pt]%
    \hbox to0pt{\verb"        {The \texttt{SCfigure} environment should be used if you"}\\[-2pt]%
    \hbox to0pt{\verb"        would like a side caption.}"}\\[-2pt]%
    \hbox to0pt{\verb"      \label{scfigure}"}\\[-2pt]%
    \hbox to0pt{\verb"  \end{SCfigure}"}%
    \hbox to0pt{\rule[-10pt]{\textwidth}{0.5pt}}%
      \caption[Figure with side caption]
        {The \texttt{SCfigure} environment should be used if you
          would like a side caption.\vspace{119pt}}
      \label{scfigure}
  \end{SCfigure}

To typeset the figure caption to the side, we recommend using the \verb"sidecap" style file; for coding see Figure~\ref{scfigure}. Note that the \verb"[50]" option on the first line will make the caption extend to the full width of the page, and is therefore essential.

Should you have a long caption which extends below the figure, you will need to revert to the \verb"figure" environment.
%

\subsection{Figures wider than the text width}

\begin{figure}
    \widefigure
    \includegraphics[width=33pc]{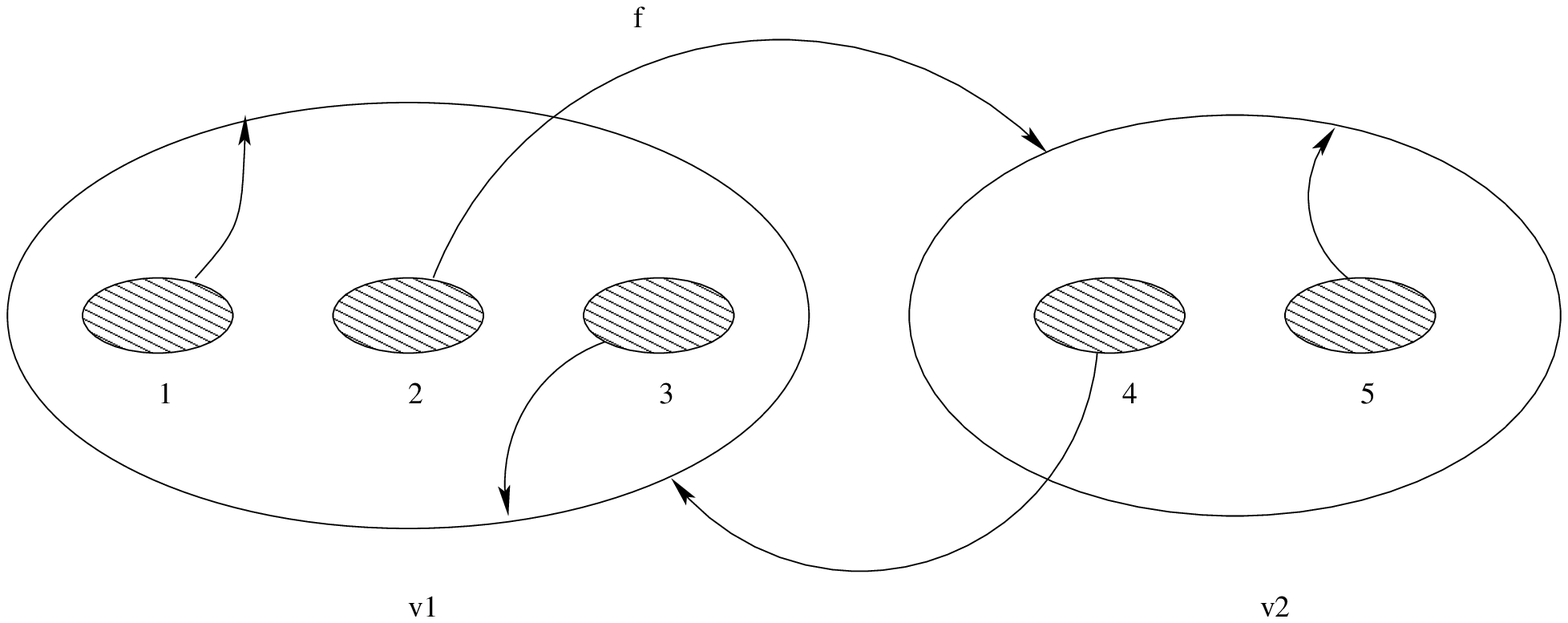}
    \caption[A wide figure]
      {Wide figures (33pc) may extend into the left-hand margin.}
    \label{widefigure}
\rule[-20pt]{\textwidth}{0.5pt}
\begin{verbatim}
  \begin{figure}
    \widefigure
    \includegraphics[width=33pc]{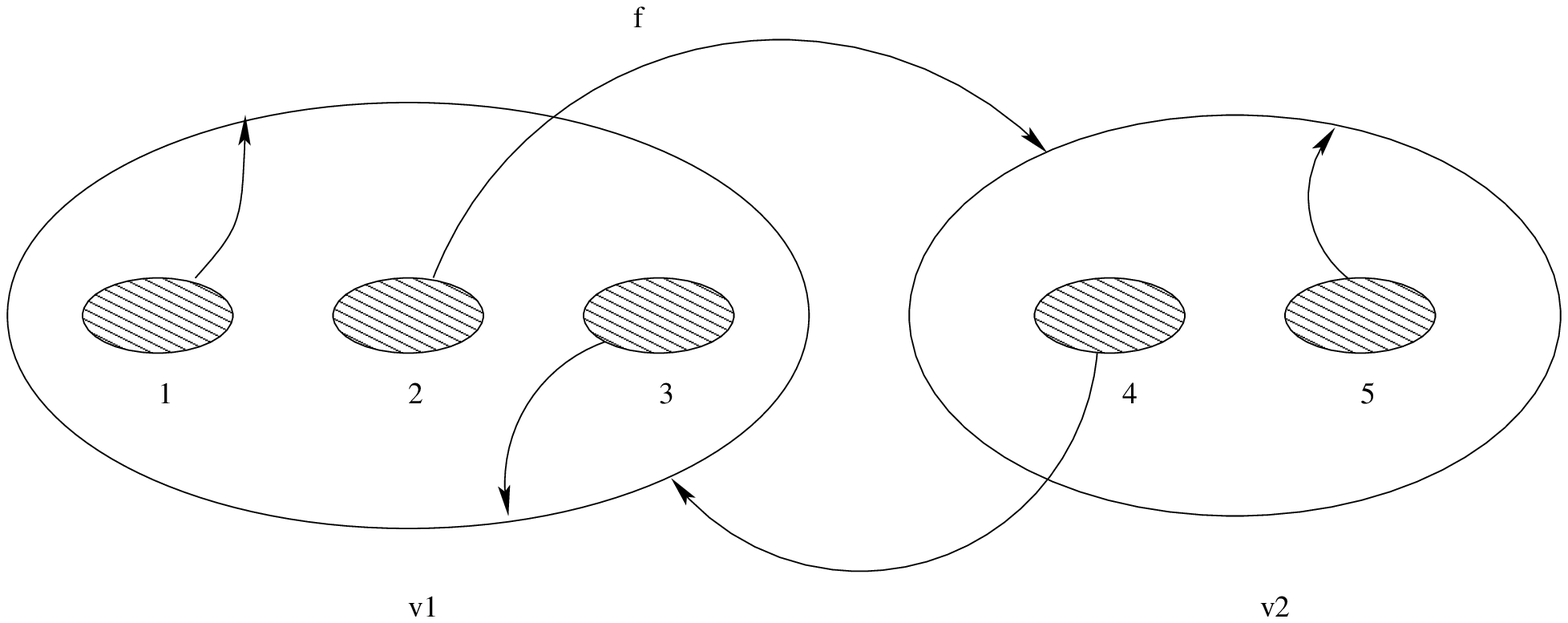}
    \caption[A wide figure]
      {Wide figures (33pc) may extend into the left-hand margin.}
    \label{widefigure}
  \end{figure}
\end{verbatim}
\rule[20pt]{\textwidth}{0.5pt}
  \end{figure}

The \cambridge\ design will allow you to have figures exactly 33pc wide; these will extend into the left margin. For these, you would use the standard \texttt{figure} environment, but you need to add \verb"\widefigure" before the graphic is included, as shown in Figure~\ref{widefigure}.

\section{Tables}

The \cambridge\ class file will cope with most positioning of your tables. Tables need to be keyed slightly differently depending on their width. When you are first writing your book, assume that they will all be the full width of the page (see Section~\ref{overtwothirds}), and adjust this later.

As captions are normally positioned above tables, the caption must be included first, then the label, then the table. Note that if you are producing a list of tables (using \verb"\listoftables"), you need to repeat the caption in square braces, but without the full point.

Note that the \verb"\begin{tabular}{@{}lllllll@{}}" command always starts and finishes with \verb"@{}". This removes the space to the left of the first column, and to the right of the last column. You need to do this as the specification for tables in the \cambridge\ design is flush-left.

\subsection{Tables between two-thirds and full text width}
\label{overtwothirds}

If the width of the table lies between two-thirds and the full text width (in other words, between 19pc and 29pc), the table may be typeset using \verb"table*" (see Table~\ref{problimits}).

The \verb"table*" environment will give full-page width \verb"\hline"s; as a result you will most likely need to increase the space between columns to make the table fill the width of the page. If you are producing the final typeset version of your book, the space may be increased by changing \verb"\tabcolsep" as shown in Table~\ref{problimits}. If not, do not concern yourself with these values as they may change with a font substitution.

  \begin{table*}
    \caption[Probability left and right for the cost data]
      {Probability left and right of the exact confidence limits
      for the ratio of exponential means for the cost data.
      The exact limits were computed using the $F$~distribution and
      the approximate probability using the $r^*$ approximation.
      In this example, the Lugannani--Rice and $r^*$ approximations
      are identical.}
    \label{problimits}
    \addtolength\tabcolsep{4pt}
    \begin{tabular}{@{}lllllll@{}}
      \hline
      Exact
        & 0.20      & 0.10       & 0.05       & 0.025      & 0.001\\
      Approx. (left)
        & 0.199\,98 & 0.099\,992 & 0.049\,997 & 0.024\,999 & 0.0001\,000\\
      Approx. (right)
        & 0.200\,03 & 0.100\,239 & 0.050\,015 & 0.025\,009 & 0.0001\,000\\
      \hline
    \end{tabular}
\rule[-20pt]{\textwidth}{0.5pt}
\begin{verbatim}
  \begin{table*}
    \caption[Probability left and right for the cost data]
      {Probability left and right of the exact confidence limits
      for the ratio of exponential means for the cost data.
      The exact limits were computed using the $F$~distribution and
      the approximate probability using the $r^*$ approximation.
      In this example, the Lugannani--Rice and $r^*$ approximations
      are identical.}
    \label{problimits}
    \addtolength\tabcolsep{4pt}% to stretch columns, if required
    \begin{tabular}{@{}lllllll@{}}
      \hline
      Exact
        & 0.20      & 0.10       & 0.05       & 0.025      & 0.001\\
      Approx. (left)
        & 0.199\,98 & 0.099\,992 & 0.049\,997 & 0.024\,999 & 0.0001\,000\\
      Approx. (right)
        & 0.200\,03 & 0.100\,239 & 0.050\,015 & 0.025\,009 & 0.0001\,000\\
      \hline
    \end{tabular}
  \end{table*}
\end{verbatim}
\rule[20pt]{\textwidth}{0.5pt}
\end{table*}

\subsection{Tables less than two-thirds of the text width}

If you have a table which takes up less than two-thirds of the text width, i.e. less than 19pc, you have two options. You may either have the caption above the table, or choose the space-saving option of placing it to the side.

\subsubsection{Caption above table}

In this case, you would use the \verb"table" environment; see Table~\ref{piexample}.

  \begin{table}
    \begin{minipage}{160pt}
    \caption[Shortened table caption for the list of tables]
      {Longer table captions have to be placed inside a minipage,
      otherwise they overhang the table rules.}
    \label{piexample}
    \addtolength\tabcolsep{2pt}
      \begin{tabular}{@{}c@{\hspace{25pt}}ccc@{}}
        \hline
        Figure\footnote{\textit{Note:} You must also use a minipage
          environment if you have footnotes.} & $hA$ & $hB$ & $hC$\\
        \hline
        1 & $\exp\left(\pi i\frac58\right)$
          & $\exp\left(\pi i\frac18\right)$ & $0$\\[3pt]
        2 & $-1$    & $\exp\left(\pi i\frac34\right)$ & $1$\\[11pt]
        3 & $-4+3i$ & $-4+3i$ & $\frac74$\\[3pt]
        4 & $-2$    & $-2$    & $\frac54 i$ \\
        \hline
      \end{tabular}
    \end{minipage}
\rule[-20pt]{\textwidth}{0.5pt}
\begin{verbatim}
  \begin{table}
    \begin{minipage}{160pt}
      %  note that the square brace option below is only required
      %  if you intend to produce a list of tables
    \caption[Shortened table caption for the list of tables]
      {Longer table captions have to be placed inside a minipage,
      otherwise they overhang the table rules.}
    \label{piexample}
    \addtolength\tabcolsep{2pt}% to stretch columns, if required
      \begin{tabular}{@{}c@{\hspace{25pt}}ccc@{}}
        \hline
        Figure\footnote{\textit{Note:} You must also use a minipage
          environment if you have footnotes.} & $hA$ & $hB$ & $hC$\\
        \hline
        1 & $\exp\left(\pi i\frac58\right)$
          & $\exp\left(\pi i\frac18\right)$ & $0$\\[3pt]
        2 & $-1$    & $\exp\left(\pi i\frac34\right)$ & $1$\\[11pt]
        3 & $-4+3i$ & $-4+3i$ & $\frac74$\\[3pt]
        4 & $-2$    & $-2$    & $\frac54 i$ \\
        \hline
      \end{tabular}
    \end{minipage}
  \end{table}
\end{verbatim}
\rule[20pt]{\textwidth}{0.5pt}
  \end{table}

\subsubsection{Caption to the side of table}

To typeset the table caption to the side, we recommend using the \verb"sidecap" style file; for coding see Table~\ref{sctable}. Note that the \verb"[50]" option on the first line will make the caption extend to the full width of the page, and is therefore essential.

Should you have a long caption which extends below the table, you will need to revert to the \verb"table" environment.

  \begin{SCtable}[50]
      \caption[Table with side caption]
          {The \texttt{SCtable} environment should be used if you
          would like a side caption. Measured and theoretical
          temperatures~($^\circ$C) in~20~sections of a reactor
          (Cox and Snell, 1981, Example~D).}
    \label{sctable}
    \begin{tabular}{@{}cc@{}}
      \hline
      Measured & Theoretical\\
      \hline
      431 & 432\\
      450 & 470\\
      431 & 442\\
      453 & 439\\
      481 & 502\\
      449 & 445\\
      441 & 455\\
      \hline
      \end{tabular}\\[13pt]
      \begin{tabular}{@{}p{0pt}@{}}
      \rule[2pt]{\textwidth}{0.5pt}\\
      \verb"  \begin{SCtable}[50]"\\
      \verb"    \caption[Table with side caption]"\\
      \verb"      {The \texttt{SCtable} environment should be used if you"\\
      \verb"      would like a side caption. Measured and theoretical"\\
      \verb"      temperatures~($^\circ$C) in~20~sections of a reactor"\\
      \verb"      (Cox and Snell, 1981, Example~D).}"\\
      \verb"    \label{sctable}"\\
      \verb"      \begin{tabular}{@{}cc@{}}"\\
      \verb"        \hline"\\
      \verb"        Measured & Theoretical\\"\\
      \verb"        \hline"\\
      \verb"        431 & 432\\"\\
      \verb"        450 & 470\\"\\
      \verb"        431 & 442\\"\\
      \verb"        453 & 439\\"\\
      \verb"        481 & 502\\"\\
      \verb"        449 & 445\\"\\
      \verb"        441 & 455\\"\\
      \verb"        \hline"\\
      \verb"      \end{tabular}"\\
      \verb"  \end{SCtable}"\\
      \rule[2pt]{\textwidth}{0.5pt}\\
    \end{tabular}
  \end{SCtable}

\subsection{My vertical rules have disappeared}

Vertical rules in tables are not \cambridge\ style, and have been automatically removed; this gives your document a truly professional look. Instead of vertical rules, we recommend the use of extra horizontal space, see Section~\ref{addhoriz}. The rules have been removed by redefining the \verb"tabular" environment. The amended definition also inserts extra vertical space above and below the horizontal rules (produced by \verb"\hline").

If you really must have them reinstated, read Section~\ref{reinstate}.

\subsection{Reinstating the vertical rules}
\label{reinstate}
Authors can revert to the standard \LaTeX\ style, if necessary. Tables will take on a rather squashed appearance, as in the \LaTeX\ book, whereby there is no added space around horizontal rules. Add the command \verb"\reinstaterules" in the preamble, and re-run your files through \LaTeX.

\subsection{There is very little space around the rules in my~table}
Tables revert to the standard, rather squashed look of standard \LaTeX\ tables for two reasons:
\begin{enumerate}
  \item you are using \verb"array.sty"; or
  \item you have chosen to reinstate vertical rules (see Section~\ref{reinstate})
\end{enumerate}
In both cases, the tabular environment is redefined.

\subsection{Adding space between columns}
\label{addhoriz}
You can add space (2pt in this example) between every column using\linebreak \verb"\addtolength\tabcolsep{2pt}". However, if you only wanted to expand the space between columns~1 and~2 to~25pt, you would do this using\linebreak \verb"\begin{tabular}{@{}c@{\hspace{25pt}}ccc@{}}" (see Table~\ref{piexample}).

\subsection{Adding space between rows}
If you need some form of separation between rows (for example, between rows~2 and~3 in the body of Table~\ref{piexample}), adding \verb"[11pt]" immediately after the double backslash at the end of row~2 will add a 11pt vertical space (the equivalent of a blank line at this typesize). This is neater than adding another horizontal line.

\section{Landscape figures and tables, using rotating.sty}

Landscape figures and tables (floats) may be typeset using the \verb"rotating.sty" package. Note that the direction of rotation depends on the page number -- which requires at least two passes through \LaTeX. If we are going to know whether pages are odd or even, we need to use the \verb"\pageref" mechanism, and labels. But labels won't work unless the user has put in a caption. \textit{Beware!}

In addition to \verb"rotating.sty", you should also include \verb"floatpag.sty" and the command \verb"\rotfloatpagestyle{empty}". This combination ensures that headers and footers are removed from the float page:
\begin{verbatim}
  \usepackage{rotating}
  \usepackage{floatpag}
  \rotfloatpagestyle{empty}
\end{verbatim}
In some DVI previewers, floats may not appear rotated. If this happens, you need to convert the DVI file to PostScript or PDF.

Occasionally, when you convert a PostScript file to a PDF file, you may find that the page comes out upside-down. There will be a setting to change this. For instance, if you are using PDFCreator 0.9.7, choose the following options in this sequence:
\begin{description}
  \item Options -- Program -- PDF -- Auto-Rotate Pages: Change to `None'.
\end{description}
Other programs will have similar procedures.

\subsection{Coding for landscape figures}

\begin{sidewaysfigure}
    \centering
    \includegraphics[scale=0.85]{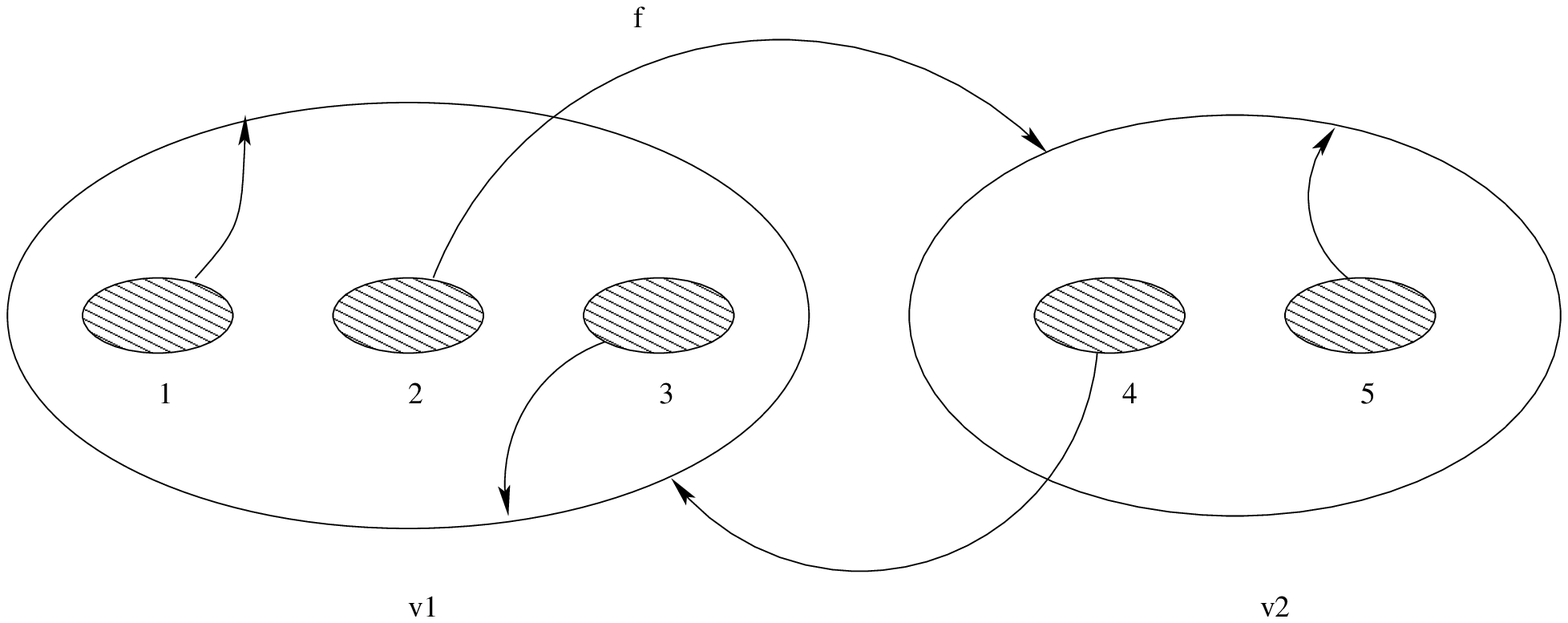}
    \caption[Landscape figure]{A Cantor repeller. Figure captions
      will be centred by default.}
    \label{sidecantor}
  \end{sidewaysfigure}

The landscape figure (Figure~\ref{sidecantor}) was typeset using the following coding:
\begin{verbatim}
  \begin{sidewaysfigure}
    \centering
    \includegraphics[scale=0.85]{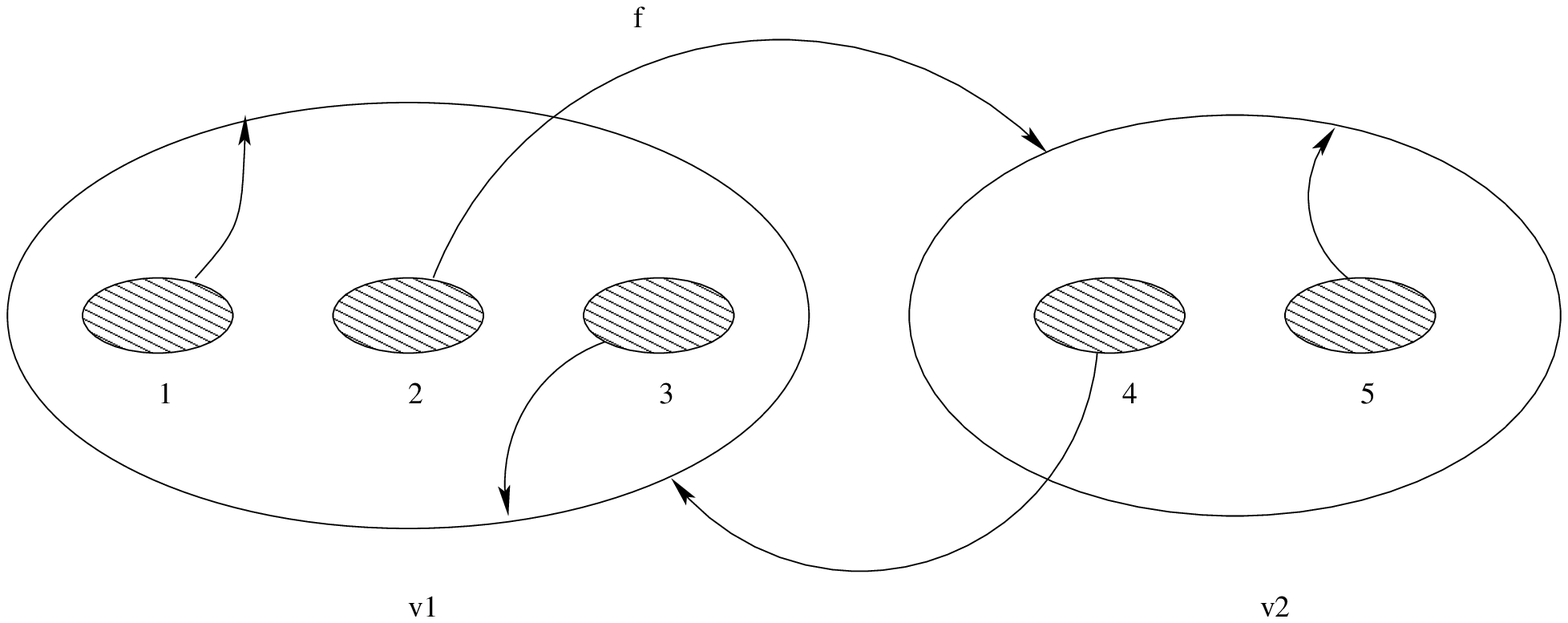}
    %  note that the square brace option below is only required
    %  if you intend to produce a list of illustrations
    \caption[Landscape figure]{A Cantor repeller. Figure captions
      will be centred by default.}
    \label{sidecantor}
  \end{sidewaysfigure}
\end{verbatim}

\subsection{Coding for landscape tables}

Table~\ref{sideways} has been produced using the following coding:
\begin{smallverbatim}
\begin{sidewaystable}
\begin{minipage}{465pt}
  \caption[Landscape table]{Grooved ware and beaker features, their finds and
    radiocarbon dates. For a breakdown of the pottery assemblages see
    Tables~I and~III; for the flints see Tables~II and~IV; for the animal
    bones see Table~V.}
  \label{sideways}
  \addtolength\tabcolsep{-2pt}
  \begin{tabular}{@{}lcccllccccc@{}}
  \hline
  Context & Length & Breadth/  & Depth & Profile & Pottery & Flint & Animal
                                                   & Stone & Other & C14 Dates\\
  && Diameter &&&&& Bones\\[5pt]
  & m & m & m\\
  \hline\\[-5pt]
  \multicolumn{10}{@{}l}{\textbf{Grooved Ware}}\\
  784 & --   & 0.9$\phantom{0}$ &0.18  & Sloping U & P1      & $\times$46
        & $\phantom{0}$$\times$8 && $\times$2 bone & 2150 $\pm$100\,\textsc{bc}\\
  785 & --   & 1.00             &0.12   & Sloping U & P2--4  & $\times$23
                                           & $\times$21 & Hammerstone & -- & --\\
  962 & --   & 1.37             &0.20   & Sloping U & P5--6  & $\times$48
                     & $\times$57 & --& --& 1990 $\pm$80\,\textsc{bc} (Layer 4)\\
  &&&&&&&&&& 1870 $\pm$90\,\textsc{bc} (Layer 1)\\
  983 & 0.83 & 0.73             &0.25   & Stepped U & --     & $\times$18
                                & $\phantom{0}$$\times$8 & -- & Fired clay & --\\
  &&&&&&&&&&\\
  \multicolumn{10}{@{}l}{\textbf{Beaker}}\\
  552 & --   & 0.68             & 0.12  & Saucer    & P7--14 & --           & --
                                                                   & -- &-- &--\\
  790 & --   & 0.60             & 0.25  & U         & P15    & $\times$12   & --
                                                      & Quartzite-lump & -- &--\\
  794 & 2.89 & 0.75             & 0.25  & Irreg.    & P16    & $\phantom{0}$$\times$3
                                                              & -- & -- &-- &--\\
  \hline
  \end{tabular}%
\end{minipage}
\end{sidewaystable}
\end{smallverbatim}
\begin{sidewaystable}
\begin{minipage}{465pt}
  \caption[Landscape table]{Grooved ware and beaker features, their finds and
    radiocarbon dates. For a breakdown of the pottery assemblages see
    Tables~I and~III; for the flints see Tables~II and~IV; for the animal
    bones see Table~V.}
  \label{sideways}
  \addtolength\tabcolsep{-2pt}
  \begin{tabular}{@{}lcccllccccc@{}}
  \hline
  Context & Length & Breadth/  & Depth & Profile & Pottery & Flint & Animal
                                                   & Stone & Other & C14 Dates\\
  && Diameter &&&&& Bones\\[5pt]
  & m & m & m\\
  \hline\\[-5pt]
  \multicolumn{10}{@{}l}{\textbf{Grooved Ware}}\\
  784 & --   & 0.9$\phantom{0}$ &0.18  & Sloping U & P1      & $\times$46
        & $\phantom{0}$$\times$8 && $\times$2 bone & 2150 $\pm$100\,\textsc{bc}\\
  785 & --   & 1.00             &0.12   & Sloping U & P2--4  & $\times$23
                                           & $\times$21 & Hammerstone & -- & --\\
  962 & --   & 1.37             &0.20   & Sloping U & P5--6  & $\times$48
                     & $\times$57 & --& --& 1990 $\pm$80\,\textsc{bc} (Layer 4)\\
  &&&&&&&&&& 1870 $\pm$90\,\textsc{bc} (Layer 1)\\
  983 & 0.83 & 0.73             &0.25   & Stepped U & --     & $\times$18
                                & $\phantom{0}$$\times$8 & -- & Fired clay & --\\
  &&&&&&&&&&\\
  \multicolumn{10}{@{}l}{\textbf{Beaker}}\\
  552 & --   & 0.68             & 0.12  & Saucer    & P7--14 & --           & --
                                                                   & -- &-- &--\\
  790 & --   & 0.60             & 0.25  & U         & P15    & $\times$12   & --
                                                      & Quartzite-lump & -- &--\\
  794 & 2.89 & 0.75             & 0.25  & Irreg.    & P16    & $\phantom{0}$$\times$3
                                                              & -- & -- &-- &--\\
  \hline
  \end{tabular}%
\end{minipage}
\end{sidewaystable}


\chapter{Mathematical solutions}
\label{mathsol}

\section{Why are we using amsthm.sty?}

Many authors are already using this style file, so we have decided that rather than re-invent the wheel, we will make it part of our distribution. This means that at the top of the root file must include the following lines:\\[0.5\baselineskip]
\verb"  \documentclass{"\texttt{\cambridge}\verb"}"\\
\verb"  \usepackage{amsmath}"\\
\verb"  \usepackage{amsthm}"\\[0.5\baselineskip]
As mentioned in Chapter~\ref{intro}, if your book does not use theorems, proofs, etc., then there is no need to include the amsthm package, but you do need to include these files to run this guide through \LaTeX. Note that if you are also using \verb"amsmath.sty", it \emph{must} precede \verb"amsthm.sty".

The instructions for amsthm.sty are documentated separately in \texttt{amsthdoc.pdf}. We are including \texttt{amsthm.sty} and \texttt{amsthdoc.pdf} in this distribution for your convenience, but you may find more recent versions on the web. The following sections discuss the basic features, plus a few extras.

To save time, you may cut and paste the code in Appendix~\ref{amsthmcommands} into your root file. This is a comprehensive (but not necessarily a complete) list of theorem-like environments you may wish to use.

The \verb"amsthm" commands used in this guide are detailed in Appendix~\ref{rootfile}. They are simply a subset of commands from Appendix~\ref{amsthmcommands}; some illustrate unnumbered versions.

Please note that theorems, lemmas, corollaries, propositions, conjectures, criteria, algorithms, definitions, conditions, problems, examples, etc.\ should be numbered in a single sequence, either by chapter (Chapter~4 would have \mbox{\textsc{\spacedheader{definition}} 4.1},
\mbox{\textsc{\spacedheader{lemma}} 4.2},
\mbox{\textsc{\spacedheader{lemma}} 4.3},
\mbox{\textsc{\spacedheader{proposition}} 4.4},
\mbox{\textsc{\spacedheader{corollary}} 4.5}) or by section (%
\mbox{\textsc{\spacedheader{definition}} 4.1.1},
\mbox{\textsc{\spacedheader{lemma}} 4.1.2},
\mbox{\textsc{\spacedheader{lemma}} 4.1.3},
\mbox{\textsc{\spacedheader{proposition}} 4.1.4},
\mbox{\textsc{\spacedheader{corollary}} 4.1.5}).

To number these elements by chapter in this guide, we have used\linebreak \verb"\newtheorem{theorem}{Theorem}[chapter]". If you prefer to have the elements numbered by section, replace \verb"[chapter]" with \verb"[section]".

\section{amsthm styles}

If no \verb"\theoremstyle" command is given, the style used will be \texttt{plain}. To specify different styles, divide your \verb"\newtheorem" commands into groups and preface each group with the appropriate \verb"\theoremstyle".

\subsection{amsthm \texttt{plain} style}

The \texttt{plain} style is normally used for theorems, lemmas, corollaries, propositions, conjectures, criteria and algorithms. The \cambridge\ style calls for these to be numbered in the same sequence. The following example resets the theorem numbers for each chapter; lemmas follow in the same sequence. For demonstration purposes only, we have requested that corollaries remain unnumbered by using the starred version:
\begin{verbatim}
  \theoremstyle{plain}% default
  \newtheorem{theorem}{Theorem}[chapter]
  \newtheorem{lemma}[theorem]{Lemma}
  \newtheorem*{corollary}{Corollary}

  \begin{theorem}
    Let the scalar function\ldots
  \end{theorem}
  \begin{lemma}[Tranah 2009]
    The first-order free surface amplitudes\ldots
  \end{lemma}
  \begin{lemma}[Normansell 2010]
    The exotic behaviours of Lagrangian\ldots
  \end{lemma}
  \begin{corollary}
    Let $G$ be the free group on\ldots
  \end{corollary}
\end{verbatim}
will produce the following output:
  \begin{theorem}
    Let the scalar function\ldots
  \end{theorem}
  \begin{lemma}[Tranah 2009]
    The first-order free surface amplitudes\ldots
  \end{lemma}
  \begin{lemma}[Normansell 2010]
    The exotic behaviours of Lagrangian\ldots
  \end{lemma}
  \begin{corollary}
    Let $G$ be the free group on\ldots
  \end{corollary}
Note that corollaries would normally be in the same numbering sequence as theorems and lemmas.

\subsection{amsthm \texttt{definition} style}
\label{amsdefn}

The \texttt{definition} style is normally used for definitions and conditions. It is \cambridge\ style to continue  with the same numbering sequence as for theorems, lemmas, etc.:
\begin{verbatim}
  \theoremstyle{definition}
  \newtheorem{definition}[theorem]{Definition}
  \newtheorem{condition}[theorem]{Condition}

  \begin{definition}
    The series above is the Green function\ldots
  \end{definition}
  \begin{condition}
    Diagrams~3 and~4 contribute to coupling constant renormalization
      of the mean field contribution\ldots
  \end{condition}
\end{verbatim}
will produce the following output:
  \begin{definition}
    The series above is the Green function\ldots
  \end{definition}
  \begin{condition}
    Diagrams~3 and~4 contribute to coupling constant renormalization
      of the mean field contribution\ldots
  \end{condition}

\subsection{amsthm \texttt{remark} style}
The \texttt{remark} style is normally used for remarks, notes, notation, claims, summary, acknowledgements, cases, conclusions. Authors are free to define their preferred numbering systems for these, but they must be separate from the theorem sequence.
\begin{verbatim}
  \theoremstyle{remark}
  \newtheorem{remark}{Remark}[chapter]
  \newtheorem{notation}[remark]{Notation}
  \newtheorem*{case}{Case}

  \begin{remark}
    The absolute amplitude of a stratified wake\ldots
  \end{remark}
  \begin{notation}
    For a poset $P$ and $\pi,\sigma \in P$ with $\pi < \sigma$
    we denote by $[\pi,\sigma]$ the interval\ldots
  \end{notation}
  \begin{case}
    The profiles of quadratic fluctuations\ldots
  \end{case}
\end{verbatim}
will produce the following output:
  \begin{remark}
    The absolute amplitude of a stratified wake\ldots
  \end{remark}
  \begin{notation}
    For a poset $P$ and $\pi,\sigma \in P$ with $\pi < \sigma$
    we denote by $[\pi,\sigma]$ the interval\ldots
  \end{notation}
  \begin{case}
    The profiles of quadratic fluctuations\ldots
  \end{case}

\section{Proofs}
\label{proofs}

The \verb"proof" environment is also part of the amsthm package, and provides a consistent format for proofs.
 For example,
\begin{verbatim}
  \begin{proof}
    Use $K_\lambda$ and $S_\lambda$ to translate combinators
    into $\lambda$-terms. For the converse, translate
    $\lambda x$ \ldots by [$x$] \ldots and use induction
    and the lemma.
  \end{proof}
\end{verbatim}
produces the following:
  \begin{proof}
    Use $K_\lambda$ and $S_\lambda$ to translate combinators
    into $\lambda$-terms. For the converse, translate
    $\lambda x$ \ldots by [$x$] \ldots and use induction
    and the lemma.
  \end{proof}

\subsection{Changing the word `Proof' to something else}

An optional argument allows you to substitute a different name for the standard `Proof'. To change the proof heading to read `Proof of the Pythagorean Theorem', key the following:
\begin{verbatim}
  \begin{proof}[Proof of the Pythagorean Theorem]
    Start with a generic right-angled triangle\ldots
  \end{proof}
\end{verbatim}
which produces:
  \begin{proof}[Proof of the Pythagorean Theorem]
    Start with a generic right-angled triangle\ldots
  \end{proof}

\subsection{Typesetting a proof without a \qedsymbol}

This is not part of the amsthm package. Use the \verb"proof*" version. For example,
\begin{verbatim}
  \begin{proof*}
    The apparent virtual mass coefficient\ldots
  \end{proof*}
\end{verbatim}
produces the following:
  \begin{proof*}
    The apparent virtual mass coefficient\ldots
  \end{proof*}

\subsection{Placing the \qedsymbol\ after a displayed equation}

To avoid the \qedsymbol\ dropping onto the following line at the end of a proof,
\begin{verbatim}
  \begin{proof}
    \ldots and, as we are all aware,
    \[
       E=mc^2. \qedhere
    \]
  \end{proof}
\end{verbatim}
produces the following:
  \begin{proof}
    \ldots and, as we are all aware,
    \[
       E=mc^2. \qedhere
    \]
  \end{proof}
When used with the amsmath package, version~2 or later, \verb"\qedhere" will position \qedsymbol\ flush right; with earlier versions, \qedsymbol\ will be spaced a quad away
from the end of the text or display.

If \verb"\qedhere" produces an error message in an equation, try using \verb"\mbox{\qedhere}" instead.

\subsection{Placing the \qedsymbol\ after a displayed eqnarray}

This is also not part of the amsthm package. To enable this, you need to used the starred version of \verb"proof", and add both \verb"\arrayqed" and \verb"\arrayqedhere", as shown in the following example:
\begin{verbatim}
  \begin{proof*}
    The following equations prove the theorem:
      \arrayqed
        \begin{eqnarray}
          \epsilon &=& -\frac{1}{2}U_0\frac{\mathrm{d}q'^2}
                       {\mathrm{d}x}\nonumber\\
                   &=& 10\nu\frac{q'^2}{\lambda^2}
        \arrayqedhere
        \end{eqnarray}
  \end{proof*}
\end{verbatim}
produces the following:
  \begin{proof*}
    The following equations prove the theorem:
      \arrayqed
        \begin{eqnarray}
          \epsilon &=& -\frac{1}{2}U_0\frac{\mathrm{d}q'^2}
                       {\mathrm{d}x}\nonumber\\
                   &=& 10\nu\frac{q'^2}{\lambda^2}
        \arrayqedhere
        \end{eqnarray}
  \end{proof*}

\section{Boxed equations}

You may highlight an individual equation using the standard \verb"\fbox" command as follows:
\begin{verbatim}
  \begin{equation}
    \fbox{$
    \ell_\textrm{c}(\alpha) = 2\alpha - \log \left( \mathrm{e}^\alpha
                                + \cdots + \mathrm{e}^{4\alpha} \right)
    $}
  \end{equation}
\end{verbatim}
  \begin{equation}
    \fbox{$
    \ell_\textrm{c}(\alpha) = 2\alpha - \log \left( \mathrm{e}^\alpha
                                + \cdots + \mathrm{e}^{4\alpha} \right)
    $}
  \end{equation}


  \part{Closing features}

\chapter{Reference and bibliography lists}

\section{Automatic lists using Bib\upshape{\TeX}}
There are three reference style options for the \cambridge\ design: Harvard (author--date), Vancouver (numbered), and IEEE (numbered); please consult with your editor as to which you should be using.

If you are using the multi-contributor option, you will get an unnumbered section heading `References', otherwise it will be an unnumbered chapter heading.

If you switch from one reference style to another, you must delete all .aux and .bbl files first, or you will get some undefined errors, or worse.

This guide has used the Harvard author--date style to produce the reference list on page~\pageref{refs}. Do not be alarmed that the log file contains several warnings such as\linebreak
\verb"LaTeX Warning: Label `MenshEst' multiply defined." These are as a result of demonstrating the three reference styles; this will not happen when you have chosen just one.

\subsection{Harvard author--date style}

\subsection*{http://www.ctan.org/tex-archive/macros/latex/contrib/harvard/}

First, call in \texttt{harvard.sty}. This style file is supplied with various bibliography styles; we recommend using the \texttt{agsm} option. The bibliography file for this guide (\texttt{\cambridge guide.tex}) is called \texttt{percolation.bib}. Place the \verb"\bibliography" command at the point where you would like the references to appear:
\begin{verbatim}
    \usepackage[agsm]{harvard}
      :
    \begin{document}
      :
  % \renewcommand{\refname}{Bibliography}
    \bibliography{percolation}
\end{verbatim}
Note that if you uncomment the third line shown above, you can change the heading from `References' to `Bibliography'. Next, \LaTeX\ your book twice. Then run \textsc{Bib}\TeX\ by executing the command\\[0.5\baselineskip]
\verb"  bibtex "\texttt{\cambridge guide}\\[0.5\baselineskip]
Finally, run your book through \LaTeX\ twice again. This series of runs will generate a file called \texttt{\cambridge guide.bbl}, which will then be included by \verb"\bibliography{percolation}".

Here are the basic citation commands available in the Harvard package; further details can be found in the documentation file \verb"harvard.pdf". Bear in mind that Menshikov (1985) or (Menshikov 1985) read best, depending on context:\\*[0.5\baselineskip]
\begin{tabular}{@{}ll@{}}
\verb"\citeasnoun{MenshEst}"
    & $\rightarrow\enskip$Menshikov (1985)\\
\verb"\citeasnoun[Appendix B]{MenshEst}"
    & $\rightarrow\enskip$Menshikov (1985, Appendix~B)\\
\verb"\cite{MenshEst}"
    & $\rightarrow\enskip$(Menshikov 1985)\\
\verb"\cite[Appendix B]{MenshEst}"
    & $\rightarrow\enskip$(Menshikov 1985, Appendix B)\\
\verb"\possessivecite{MenshEst}"
    & $\rightarrow\enskip$Menshikov's (1985)\\
\verb"\citeaffixed{MenshEst,Reimer}{e.g.}"
    & $\rightarrow\enskip$(e.g. Menshikov 1985, Reimer 2000)\\
\verb"\citeyear*{MenshEst,Reimer}"
    & $\rightarrow\enskip$1985, 2000\\
\verb"\citeyear{MenshEst,Reimer}"
    & $\rightarrow\enskip$(1985, 2000)\\
\verb"\citename{MenshEst}"
    & $\rightarrow\enskip$Menshikov
\end{tabular}\\[0.5\baselineskip]
\noindent Suppose you have cited 8 entries from the `percolation' database, e.g. \verb"\cite{MenshEst}"; \verb"\cite{Kasymp}"; \verb"\cite{Reimer}"; \verb"\cite{HamMaz94}"; \verb"\cite{HamLower}"; \verb"\cite{AiBar87}"; \verb"\cite{MMS}"; and \verb"\cite{HamAtomBond}"; the output will be just those 8~citations; see below.

\subsection*{Output from harvard author--date style}
\begin{harvardoutput}
\item Aizenman, M. \&\ Barsky, D.~J. (1987), `Sharpness of the phase transition in percolation models', {\em Comm. Math. Phys.} \textbf{108},~489--526.

\item Hammersley, J.~M. (1957), `Percolation processes: Lower bounds for the critical probability', {\em Ann. Math. Statist.} \textbf{28},~790--795.

\item Hammersley, J.~M. (1961), `Comparison of atom and bond percolation processes', {\em J. Mathematical Phys.} \textbf{2},~728--733.

\item Hammersley, J.~M. \&\ Mazzarino, G. (1994), `Properties of large Eden clusters in the plane', {\em Combin. Probab. Comput.} \textbf{3},~471--505.

\item Kesten, H. (1990), Asymptotics in high dimensions for percolation, {\em in} G.~R. Grimmett \&\ D.~J.~A. Welsh, eds, `Disorder in Physical Systems: A Volume in Honour of John Hammersley', Oxford University Press, pp.~219--240.

\item Menshikov, M.~V. (1985), `Estimates for percolation thresholds for lattices in $\textbf{R}^n$', {\em Dokl. Akad. Nauk SSSR} \textbf{284},~36--39.

\item Menshikov, M.~V., Molchanov, S.~A. \&\ Sidorenko, A.~F. (1986), Percolation theory and some applications, {\em in} `Probability theory. Mathematical statistics. Theoretical cybernetics, Vol. 24 (Russian)', Akad. Nauk SSSR Vsesoyuz. Inst. Nauchn. i Tekhn. Inform., pp.~53--110. Translated in {\em J. Soviet Math}. \textbf{42} (1988), no. 4, 1766--1810.

\item Reimer, D. (2000), `Proof of the van den Berg--Kesten conjecture', {\em Combin. Probab. Comput.} \textbf{9},~27--32.

\end{harvardoutput}

\subsection*{Harvard author--date style -- keying in your own reference list}
You do not have to use \textsc{Bib}\TeX\ to generate your list of references; the above list may be keyed as follows:
\begin{verbatim}
\begin{harvardoutput}
\item Aizenman, M. \&\ Barsky, D.~J. (1987), `Sharpness...~489--526.
\item Hammersley, J.~M. (1957), `Percolation...~790--795.
\item Hammersley, J.~M. (1961), `Comparison of atom...~728--733.
\item Hammersley, J.~M. \&\ Mazzarino, G. (1994), `Properties...~471--505.
\item Kesten, H. (1990), Asymptotics in high dimensions...~219--240.
\item Menshikov, M.~V. (1985), `Estimates for percolation...~36--39.
\item Menshikov, M.~V., Molchanov, S.~A. \&\ Sidorenko, A.~F....1766--1810.
\item Reimer, D. (2000), `Proof of the van den Berg--Kesten...~27--32.
\end{harvardoutput}
\end{verbatim}

\subsection{Vancouver numbered style}

\subsection*{http://www.ctan.org/tex-archive/biblio/bibtex/contrib/vancouver/}

First, call in the vancouver bibliography style file (\verb"vancouver.bst") as shown below. The bibliography file for this guide (\texttt{\cambridge guide.tex}) is called \texttt{percolation.bib}. Place the \verb"\bibliography" command at the point where you would like the references to appear:
\begin{verbatim}
  % \removesquarebraces
      :
    \begin{document}
      :
    \bibliographystyle{vancouver}
      :
  % \renewcommand{\refname}{Bibliography}
    \bibliography{percolation}
\end{verbatim}
Note that if you uncomment the first line, \verb"\removesquarebraces", the square braces will be removed from the final listing (but will remain in place for citations). If you uncomment the fourth line shown above, you can change the heading from `References' to `Bibliography'. Next, \LaTeX\ your book twice. Then run \textsc{Bib}\TeX\ by executing the command\\[0.5\baselineskip]
\verb"  bibtex "\texttt{\cambridge guide}\\[0.5\baselineskip]
Finally, run your book through \LaTeX\ twice again. This series of runs will generate a file called \texttt{\cambridge guide.bbl}, which will then be included by \verb"\bibliography{percolation}".

Here are the basic citation commands available in the Vancouver package; further details can be found in the documentation file \verb"vancouver.pdf". Note that you may have more than one entry within the \verb"\cite" command:\\*[0.5\baselineskip]
\begin{tabular}{@{}ll@{}}
\verb"\cite{MenshEst}"
    & $\rightarrow\enskip$[1]\\
\verb"\cite{MenshEst,Reimer}"
    & $\rightarrow\enskip$[1, 3]\\
\verb"\cite[Chapter~2]{MenshEst}"
    & $\rightarrow\enskip$[1, Chapter~2]\\
\end{tabular}\\[0.5\baselineskip]
\noindent Suppose you have cited 10 entries from the `percolation' database, e.g. \verb"\cite{MenshEst}"; \verb"\cite{Kasymp}"; \verb"\cite{Reimer}"; \verb"\cite{HamMaz94}"; \verb"\cite{HamLower}"; \verb"\cite{AiBar87}"; \verb"\cite{MMS}"; \verb"\cite{HamAtomBond}";  \verb"\cite{HamMaz83}" and \verb"\cite{HamWelsh}"; the output will be just those 10~citations; see below.

\subsection*{Output from vancouver numbered style}
\begin{vancouveroutput}{10}

\bibitem{MenshEst}
Menshikov MV.
\newblock Estimates for percolation thresholds for lattices in {${\bf R}\sp
  n$}.
\newblock Dokl Akad Nauk SSSR. 1985;284:36--39.

\bibitem{Kasymp}
Kesten H.
\newblock Asymptotics in high dimensions for percolation.
\newblock In: Grimmett GR, Welsh DJA, editors. Disorder in Physical Systems: A
  Volume in Honour of John Hammersley. Oxford University Press; 1990. p.
  219--240.

\bibitem{Reimer}
Reimer D.
\newblock Proof of the van den {B}erg--{K}esten conjecture.
\newblock Combin Probab Comput. 2000;9:27--32.

\bibitem{HamMaz94}
Hammersley JM, Mazzarino G.
\newblock Properties of large {E}den clusters in the plane.
\newblock Combin Probab Comput. 1994;3:471--505.

\bibitem{HamLower}
Hammersley JM.
\newblock Percolation processes: {L}ower bounds for the critical probability.
\newblock Ann Math Statist. 1957;28:790--795.

\bibitem{AiBar87}
Aizenman M, Barsky DJ.
\newblock Sharpness of the phase transition in percolation models.
\newblock Comm Math Phys. 1987;108:489--526.

\bibitem{MMS}
Menshikov MV, Molchanov SA, Sidorenko AF.
\newblock Percolation theory and some applications.
\newblock In: Probability theory. Mathematical statistics. Theoretical
  cybernetics, Vol. 24 (Russian). Akad. Nauk SSSR Vsesoyuz. Inst. Nauchn. i
  Tekhn. Inform.; 1986. p. 53--110.
\newblock Translated in {\em J. Soviet Math}. {\bf 42} (1988), no. 4,
  1766--1810.

\bibitem{HamAtomBond}
Hammersley JM.
\newblock Comparison of atom and bond percolation processes.
\newblock J Mathematical Phys. 1961;2:728--733.

\bibitem{HamMaz83}
Hammersley JM, Mazzarino G.
\newblock Markov fields, correlated percolation, and the {I}sing model.
\newblock In: The mathematics and physics of disordered media (Minneapolis,
  Minn., 1983). vol. 1035 of Lecture Notes in Math. Springer; 1983. p.
  201--245.

\bibitem{HamWelsh}
Hammersley JM, Welsh DJA.
\newblock First-passage percolation, subadditive processes, stochastic
  networks, and generalized renewal theory.
\newblock In: Proc. Internat. Res. Semin., Statist. Lab., Univ. California,
  Berkeley, Calif. Springer; 1965. p. 61--110.

\end{vancouveroutput}

\subsection*{Vancouver numbered style -- keying in your own reference list}
You do not have to use \textsc{Bib}\TeX\ to generate your list of references; the above list may be keyed as follows. Note that you need to specify the number of references (10~in this case) so that \LaTeX\ can work out how wide the margin needs to be.
\begin{verbatim}
\begin{vancouveroutput}{10}
\bibitem{} Menshikov MV. Estimates for percolation...1985;284:36--39.
\bibitem{} Kesten H. Asymptotics in high dimensions...1990. p.~219--240.
\bibitem{} Reimer D. Proof of the van den Berg--Kesten...2000;9:27--32.
\bibitem{} Hammersley JM, Mazzarino G. Properties...1994;3:471--505.
\bibitem{} Hammersley JM. Percolation processes:...1957;28:790--795.
\bibitem{} Aizenman M, Barsky DJ. Sharpness of the phase...1987;108:489--526.
\bibitem{} Menshikov MV, Molchanov SA, Sidorenko AF. Percolation...1766--1810.
\bibitem{} Hammersley JM. Comparison of atom and bond...1961;2:728--733.
\bibitem{} Hammersley JM, Mazzarino G. Markov fields,...p.~201--245.
\bibitem{} Hammersley JM, Welsh DJA. First-passage percolation,...p.~61--110.
\end{vancouveroutput}
\end{verbatim}

\subsection{IEEE numbered style}

\subsection*{http://www.ctan.org/tex-archive/macros/latex/contrib/IEEEtran/bibtex/}

First, call in the IEEE bibliography style file (IEEEtran.bst) as shown below. The bibliography file for this guide (\texttt{\cambridge guide.tex}) is called \texttt{percolation.bib}. Place the \verb"\bibliography" command at the point where you would like the references to appear:
\begin{verbatim}
  % \removesquarebraces
      :
    \begin{document}
      :
    \bibliographystyle{IEEEtran}
      :
  % \renewcommand{\refname}{Bibliography}
    \bibliography{percolation}
\end{verbatim}
Note that if you uncomment the first line, \verb"\removesquarebraces", the square braces will be removed from the final listing (but will remain in place for citations). If you uncomment the fourth line shown above, you can change the heading from `References' to `Bibliography'. Next, \LaTeX\ your book twice. Then run \textsc{Bib}\TeX\ by executing the command\\[0.5\baselineskip]
\verb"  bibtex "\texttt{\cambridge guide}\\[0.5\baselineskip]
Finally, run your book through \LaTeX\ twice again. This series of runs will generate a file called \texttt{\cambridge guide.bbl}, which will then be included by \verb"\bibliography{percolation}".

Here are the basic citation commands available in the IEEEtran package; further details can be found in the documentation file \verb"IEEEtran_bst_HOWTO.pdf". Note that you may have more than one entry within the \verb"\cite" command:\\*[0.5\baselineskip]
\begin{tabular}{@{}ll@{}}
\verb"\cite{MenshEst}"
    & $\rightarrow\enskip$[1]\\
\verb"\cite{MenshEst,Reimer}"
    & $\rightarrow\enskip$[1, 3]\\
\verb"\cite[Chapter~2]{MenshEst}"
    & $\rightarrow\enskip$[1, Chapter~2]\\
\end{tabular}\\[0.5\baselineskip]
\noindent Suppose you have cited 10 entries from the `percolation' database, e.g. \verb"\cite{MenshEst}"; \verb"\cite{Kasymp}"; \verb"\cite{Reimer}"; \verb"\cite{HamMaz94}"; \verb"\cite{HamLower}"; \verb"\cite{AiBar87}"; \verb"\cite{MMS}"; \verb"\cite{HamAtomBond}";  \verb"\cite{HamMaz83}" and \verb"\cite{HamWelsh}"; the output will be just those 10~citations; see below.

\subsection*{Output from IEEEtran numbered style}
\begin{IEEEtranoutput}{10}

\bibitem{MenshEst}
M.~V. Menshikov, ``Estimates for percolation thresholds for lattices in {${\bf
  R}\sp n$},'' \emph{Dokl. Akad. Nauk SSSR}, vol. 284, pp. 36--39, 1985.

\bibitem{Kasymp}
H.~Kesten, ``Asymptotics in high dimensions for percolation,'' in
  \emph{Disorder in Physical Systems: A Volume in Honour of John Hammersley},
  G.~R. Grimmett and D.~J.~A. Welsh, Eds.\hskip 1em plus 0.5em minus
  0.4em\relax Oxford University Press, 1990, pp. 219--240.

\bibitem{Reimer}
D.~Reimer, ``Proof of the van den {B}erg--{K}esten conjecture,'' \emph{Combin.
  Probab. Comput.}, vol.~9, pp. 27--32, 2000.

\bibitem{HamMaz94}
J.~M. Hammersley and G.~Mazzarino, ``Properties of large {E}den clusters in the
  plane,'' \emph{Combin. Probab. Comput.}, vol.~3, pp. 471--505, 1994.

\bibitem{HamLower}
J.~M. Hammersley, ``Percolation processes: {L}ower bounds for the critical
  probability,'' \emph{Ann. Math. Statist.}, vol.~28, pp. 790--795, 1957.

\bibitem{AiBar87}
M.~Aizenman and D.~J. Barsky, ``Sharpness of the phase transition in
  percolation models,'' \emph{Comm. Math. Phys.}, vol. 108, pp. 489--526, 1987.

\bibitem{MMS}
M.~V. Menshikov, S.~A. Molchanov, and A.~F. Sidorenko, ``Percolation theory and
  some applications,'' in \emph{Probability theory. Mathematical statistics.
  Theoretical cybernetics, Vol. 24 (Russian)}.\hskip 1em plus 0.5em minus
  0.4em\relax Akad. Nauk SSSR Vsesoyuz. Inst. Nauchn. i Tekhn. Inform., 1986,
  pp. 53--110, translated in {\em J. Soviet Math}. {\bf 42} (1988), no. 4,
  1766--1810.

\bibitem{HamAtomBond}
J.~M. Hammersley, ``Comparison of atom and bond percolation processes,''
  \emph{J. Mathematical Phys.}, vol.~2, pp. 728--733, 1961.

\bibitem{HamMaz83}
J.~M. Hammersley and G.~Mazzarino, ``Markov fields, correlated percolation, and
  the {I}sing model,'' in \emph{The mathematics and physics of disordered media
  (Minneapolis, Minn., 1983)}, ser. Lecture Notes in Math.\hskip 1em plus 0.5em
  minus 0.4em\relax Springer, 1983, vol. 1035, pp. 201--245.

\bibitem{HamWelsh}
J.~M. Hammersley and D.~J.~A. Welsh, ``First-passage percolation, subadditive
  processes, stochastic networks, and generalized renewal theory,'' in
  \emph{Proc. Internat. Res. Semin., Statist. Lab., Univ. California, Berkeley,
  Calif.}\hskip 1em plus 0.5em minus 0.4em\relax Springer, 1965, pp. 61--110.

\end{IEEEtranoutput}

\subsection*{IEEEtran numbered style -- keying in your own reference list}
You do not have to use \textsc{Bib}\TeX\ to generate your list of references; the above list may be keyed as follows. Note that you need to specify the number of references (10~in this case) so that \LaTeX\ can work out how wide the margin needs to be.
\begin{verbatim}
\begin{IEEEtranoutput}{10}
\bibitem{} M.~V. Menshikov, ``Estimates for percolation...pp.~36--39, 1985.
\bibitem{} H.~Kesten, ``Asymptotics in high dimensions for...pp.~219--240.
\bibitem{} D.~Reimer, ``Proof of the van den Berg--Kesten...pp.~27--32, 2000.
\bibitem{} J.~M. Hammersley and G.~Mazzarino, ``Properties...pp.~471--505, 1994.
\bibitem{} J.~M. Hammersley, ``Percolation processes: Lower...pp.~790--795, 1957.
\bibitem{} M.~Aizenman and D.~J. Barsky, ``Sharpness of the...pp.~489--526, 1987.
\bibitem{} M.~V. Menshikov, S.~A. Molchanov, and A.~F. Sidorenko,...no.~4, 1766--1810.
\bibitem{} J.~M. Hammersley, ``Comparison of atom and bond...pp.~728--733, 1961.
\bibitem{} J.~M. Hammersley and G.~Mazzarino, ``Markov fields,...pp.~201--245.
\bibitem{} J.~M. Hammersley and D.~J.~A. Welsh, ``First-passage...pp.~61--110.
\end{IEEEtranoutput}
\end{verbatim}

\nocite{MenshEst}
\nocite{Kasymp}
\nocite{Reimer}
\nocite{HamMaz94}
\nocite{HamLower}
\nocite{AiBar87}
\nocite{MMS}
\nocite{HamAtomBond}
\nocite{HamMaz83}
\nocite{HamWelsh}


\chapter{Indexes}
\label{indexes}

\section{Creating a single index using makeidx.sty}
To generate a single index, normally a subject index, the commands would take the form:
\begin{verbatim}
  \index{diffraction}
  \index{force!hydrodynamic}
  \index{force!interactive}
\end{verbatim}
The following commands are then required in the preamble:
\begin{verbatim}
  \usepackage{makeidx}
  \makeindex
\end{verbatim}
and at the point you wish your index to appear,
\begin{verbatim}
  \printindex
\end{verbatim}
Run your book through \LaTeX\ enough times so that the labels, etc., are stable. Then execute the command:\\[0.5\baselineskip]
\verb"  makeindex "\texttt{\cambridge guide}\\[0.5\baselineskip]
To include the index, you need to run \LaTeX\ one more time.

\section{Creating multiple indexes using multind.sty}
This guide has been prepared using \verb"multind.sty". This style file redefines the \verb"\makeindex", \verb"\index" and \verb"\printindex" commands to deal with multiple indexes.

Suppose you want to create an author index and a subject index. The entries should be in the text as usual, but take the following form:
\begin{verbatim}
  \index{authors}{Young, P.D.F.}
  \index{authors}{Tranah, D.A.}
  \index{authors}{Peterson, K.}
  \index{subject}{diffraction}
  \index{subject}{force!hydrodynamic}
  \index{subject}{force!interactive}
\end{verbatim}
  \index{authors}{Young, P.D.F.}%
  \index{authors}{Tranah, D.A.}%
  \index{authors}{Peterson, K.}%
  \index{subject}{diffraction}%
  \index{subject}{force!hydrodynamic}%
  \index{subject}{force!interactive}%
In the preamble, you need to add the following lines:
\begin{verbatim}
  \usepackage{multind}\ProvidesPackage{multind}
  \makeindex{authors}
  \makeindex{subject}
\end{verbatim}
It is crucial to add the command \verb"\ProvidesPackage{multind}"; this will send a message to the class file to re-style the index into the \cambridge\ style. You will get a warning in your log file:
\begin{verbatim}
  LaTeX Warning: You have requested package `',
                 but the package provides `multind'.
\end{verbatim}
which can be ignored. At the point where you wish your indexes to appear, you then need the commands:
\begin{verbatim}
  \printindex{authors}{Author index}
  \printindex{subject}{Subject index}
\end{verbatim}
Run your book through \LaTeX\ enough times so that the labels, etc., are stable. Then execute the commands:
\begin{verbatim}
  makeindex authors
  makeindex subject
\end{verbatim}
To include the indexes, you need to run \LaTeX\ one more time.

\section{Creating multiple indexes using index.sty}

This style file allows you to define new indexes. Suppose you want to create an author index as well as a normal subject index. The entries should be in the text as usual, but take the following form:
\begin{verbatim}
  \index[aut]{Young, P.D.F.}
  \index[aut]{Tranah, D.A.}
  \index[aut]{Peterson, K.}
  \index{diffraction}
  \index{force!hydrodynamic}
  \index{force!interactive}
\end{verbatim}
To create the extra author index, you need to have the following lines in the preamble:
\begin{verbatim}
  \usepackage{index}
  \makeindex
  \newindex{aut}{adx}{and}{Author index}
\end{verbatim}
At the point where you wish your indexes to appear, use:
\begin{verbatim}
  \printindex[aut]
  \printindex
\end{verbatim}
Run your book through \LaTeX\ enough times so that the labels, etc., are stable. Then execute the commands:\\[0.5\baselineskip]
\verb"  makeindex -o "\texttt{\cambridge guide.and \cambridge guide.adx}\\
\verb"  makeindex "\texttt{\cambridge guide}\\[0.5\baselineskip]
To include the indexes, you need to run \LaTeX\ one more time.

\subsection{Caution -- from the authors of index.sty}

In order to implement \verb"index.sty", it's been necessary to modify a number of \LaTeX\ commands seemingly unrelated to indexing, namely, \verb"\@starttoc", \verb"\raggedbottom", \verb"\flushbottom", \verb"\addcontents", \verb"\markboth", and \verb"\markright". Naturally, this could cause incompatibilities between \texttt{index.sty} and any style files that either redefine these same commands or make specific assumptions about how they operate.

The redefinition of \verb"\@starttoc" is particularly bad, since it introduces an incompatibility with the AMS document classes. This will be addressed soon.

In the current implementation, \texttt{index.sty} uses one output stream for each index.  Since there are a limited number of output indexes, this means that there is a limit on the number of indexes you can have in a document.  There is more information on this in \verb"index.dtx" which is part of the \verb"index.sty" distribution.\\[\baselineskip]
\textit{For these reasons, whilst all care has been taken to deal with these changes in \cambridge.cls, if you do find incompatibilities with other files, please contact us at texline@cambridge.org with your source files, class and style files, and log file.}


  \backmatter
  \appendix

\chapter{Typesetting appendices}

\section{Single-contributor books}
\subsection{How to typeset one appendix}
If you have just one appendix, say \verb"appendix.tex", you will want to generate a chapter head `Appendix' rather than `Appendix A'. Use \verb"\oneappendix" in the main file, as follows:
\begin{verbatim}
  \oneappendix
  \include{appendix}
\end{verbatim}

\subsection{How to typeset several appendices}
The coding used to generate the appendices in this guide is as follows:
\begin{verbatim}
  \appendix
  \include{appendixA}
  \include{appendixB}
  \include{appendixC}
\end{verbatim}

\section{Multi-contributor books}

\subsection{How to typeset one appendix}
If you have just one appendix, it will be the next section head and you should include the following code at the end of your chapter:
\begin{verbatim}
  \oneappendix
  \section{Appendix heading}
  \subsection{Subheading}
  \endappendix
\end{verbatim}
You will need to add \verb"\endappendix" if you have further section heads in this chapter.

\subsection{How to typeset several appendices}
The following code will genenerate Appendix~A and Appendix~B at the end of your chapter:
\begin{verbatim}
  \appendix
  \section{Appendix heading}
  \subsection{Subheading}
    :
  \section{Next appendix heading}
  \subsection{Next subheading}
  \endappendix
\end{verbatim}
Again, you will need to add \verb"\endappendix" if you have further section heads in this chapter.

\section{Numbering systems}

Equations in appendices will be numbered as follows:
\begin{equation}
  E=mc^2,
\end{equation}
and figure captions as follows:
\begin{figure}[h]
\caption[Similarity solutions]{Similarity solutions.}
\end{figure}


\chapter{amsthm commands}
\label{amsthmcommands}

The following code may be cut and pasted into your root file. Assuming you have included \verb"amsthm.sty", it will number your theorems, definitions, etc. in the same numbering sequence and by chapter, e.g.~%
\mbox{\textsc{\spacedheader{definition}} 4.1},
\mbox{\textsc{\spacedheader{lemma}} 4.2},
\mbox{\textsc{\spacedheader{lemma}} 4.3},
\mbox{\textsc{\spacedheader{proposition}} 4.4},
\mbox{\textsc{\spacedheader{corollary}} 4.5}.

If you prefer to have the elements numbered by section, e.g.~%
\mbox{\textsc{\spacedheader{definition}} 4.1.1},
\mbox{\textsc{\spacedheader{lemma}} 4.1.2},
\mbox{\textsc{\spacedheader{lemma}} 4.1.3},
\mbox{\textsc{\spacedheader{proposition}} 4.1.4},
\mbox{\textsc{\spacedheader{corollary}} 4.1.5}, replace \verb"[chapter]" on line 2 with \verb"[section]".

\begin{smallverbatim}

  \theoremstyle{plain}
  \newtheorem{theorem}{Theorem}[chapter]
  \newtheorem{lemma}[theorem]{Lemma}
  \newtheorem{corollary}[theorem]{Corollary}
  \newtheorem{proposition}[theorem]{Proposition}
  \newtheorem{conjecture}[theorem]{Conjecture}
  \newtheorem{criterion}[theorem]{Criterion}
  \newtheorem{algorithm}[theorem]{Algorithm}

  \theoremstyle{definition}
  \newtheorem{definition}[theorem]{Definition}
  \newtheorem{condition}[theorem]{Condition}

  \theoremstyle{remark}
  \newtheorem{remark}{Remark}[chapter]
  \newtheorem{note}[remark]{Note}
  \newtheorem{notation}[remark]{Notation}
  \newtheorem{claim}[remark]{Claim}
  \newtheorem{summary}[remark]{Summary}
  \newtheorem{acknowledgement}[remark]{Acknowledgement}
  \newtheorem{case}[remark]{Case}
  \newtheorem{conclusion}[remark]{Conclusion}
\end{smallverbatim}

  \include{appendixC}
  \endappendix

  \addtocontents{toc}{\vspace{\baselineskip}}
  \theendnotes


  \bibliography{percolation}\label{refs}

  \cleardoublepage



  \printindex{authors}{Author index}
  \printindex{subject}{Subject index}
